\documentclass[useAMS,usenatbib]{mn2e}
\newcommand{\beq}{\begin{equation}}
\newcommand{\eeq}{\end{equation}}
\newcommand{\beqa}{\begin{eqnarray}}
\newcommand{\eeqa}{\end{eqnarray}}
\usepackage{float}
\usepackage{graphicx}
\usepackage{amssymb}
\usepackage{amsmath}
\usepackage{datetime}
\usepackage[normalem]{ulem}
\usepackage{times}
\usepackage[export]{adjustbox}
\usepackage{soul}

\usepackage{color}
\usepackage{soul}

\title[Breakout from centrifugal magnetospheres]{
How the breakout-limited mass in B-star centrifugal magnetospheres controls their circumstellar H$\alpha$ emission}
\author[S.\ Owocki et al.]{
Stanley P.\ Owocki$^{1,2}$\thanks{email: owocki@udel.edu},
Matt E. Shultz$^1$, 
Asif ud-Doula$^3$,
Jon O.\ Sundqvist$^4$,
\newauthor
 Richard H.D.\ Townsend$^5$
and
Steven R.\ Cranmer$^6$
\\
$^1$
Departmenst of Physics \& Astronomy, 
University of Delaware, Newark, DE 19716 USA 
\\
$^2$
Bartol Research Insitute, 
University of Delaware, Newark, DE 19716 USA 
\\
$^3$ Penn State Scranton, 120 Ridge View Dr., Dunmore, PA 18512, USA.\\
$^4$ KU Leuven, Instituut voor Sterrenkunde, Celestijnenlaan 200D, 3001 Leuven, Belgium \\
$^5$ Department of Astronomy, University of Wisconsin-Madison, Madison, WI 53706, USA \\
$^6$ Department of Astrophysical and Planetary Sciences, Laboratory for Atmospheric and Space Physics, University of Colorado,  Boulder, CO 80309, USA
}

\begin{document}

\include{aas_macros}

\date{Accepted ?.  Received ?; in original form ?}

\maketitle

\label{firstpage}

\begin{abstract}
Strongly magnetic B-type stars with moderately rapid rotation form `centrifugal magnetospheres' (CMs), from the magnetic trapping of stellar wind material in a region above the Kepler co-rotation radius.
A longstanding question is whether the eventual loss of such trapped material occurs from gradual drift and/or diffusive leakage, or through sporadic `{\em centrifugal break out}'  (CBO) events, wherein magnetic tension can no longer contain the built-up mass.
We argue here that recent empirical results for Balmer-$\alpha$ emission from such B-star CMs strongly favor the CBO mechanism.
Most notably, the fact that the onset of such emission depends mainly on the field strength at the Kepler radius, and is largely {\em independent} of the stellar luminosity,
strongly disfavors any drift/diffusion process, for which the net mass balance would depend on the luminosity-dependent wind feeding rate.
In contrast, we show that in a CBO model the {\em maximum confined mass} in the magnetosphere is independent of this wind feeding rate, and 
has a dependence on field strength and Kepler radius that naturally explains the empirical scalings for the onset of H$\alpha$ 
emission, its associated equivalent width, and even its line profile shapes. 
However, the general lack of observed Balmer emission in late-B and A-type stars could  still be attributed to a residual level of diffusive or drift leakage 
that does not allow their much weaker winds to fill their CMs to the breakout level needed for such emission;
alternatively this might result from a transition to a metal-ion wind that lacks the requisite Hydrogen.
\end{abstract}


\today

\section{Introduction}

Hot luminous, massive stars of spectral type O and B have dense, high-speed, radiatively driven stellar winds
\citep{Castor75}.
In the subset ($\sim$10\%) of massive stars with strong ($> 100$G), globally ordered (often significantly dipolar) magnetic fields
\citep{Petit13}, the trapping of this wind outflow by closed magnetic loops leads to the formation of a circumstellar {\em magnetosphere},
as first discovered and characterised in $\sigma$~Ori~E by \citet{Landstreet78}.
Because of the angular momentum loss associated with 
their relatively strong, magnetised wind 
\citep{udDoula09}, magnetic O-type stars are typically
slow rotators,  with trapped wind material falling back on a dynamical timescale, giving then what's known as a ``dynamical magnetosphere" (DM).
But in magnetic B-type stars, the relatively weak stellar winds imply longer spin-down times, and so a significant fraction that still retain a moderately rapid rotation;
in cases that the associated Keplerian corotation radius $R_{\rm K}$ lies within the Alfv\'{e}n radius $R_{\rm A}$ that characterises the maximum height of closed loops, the rotational 
support leads to formation of a ``{\em centrifugal magnetosphere}'' (CM), wherein the trapped wind material accumulates into a relatively dense, stable and long-lived
`rigidly rotating magnetosphere' (RRM) 
\citep[][hereafter TO05]{Townsend05}.

\begin{figure}
\begin{center}
\includegraphics[scale=.35]{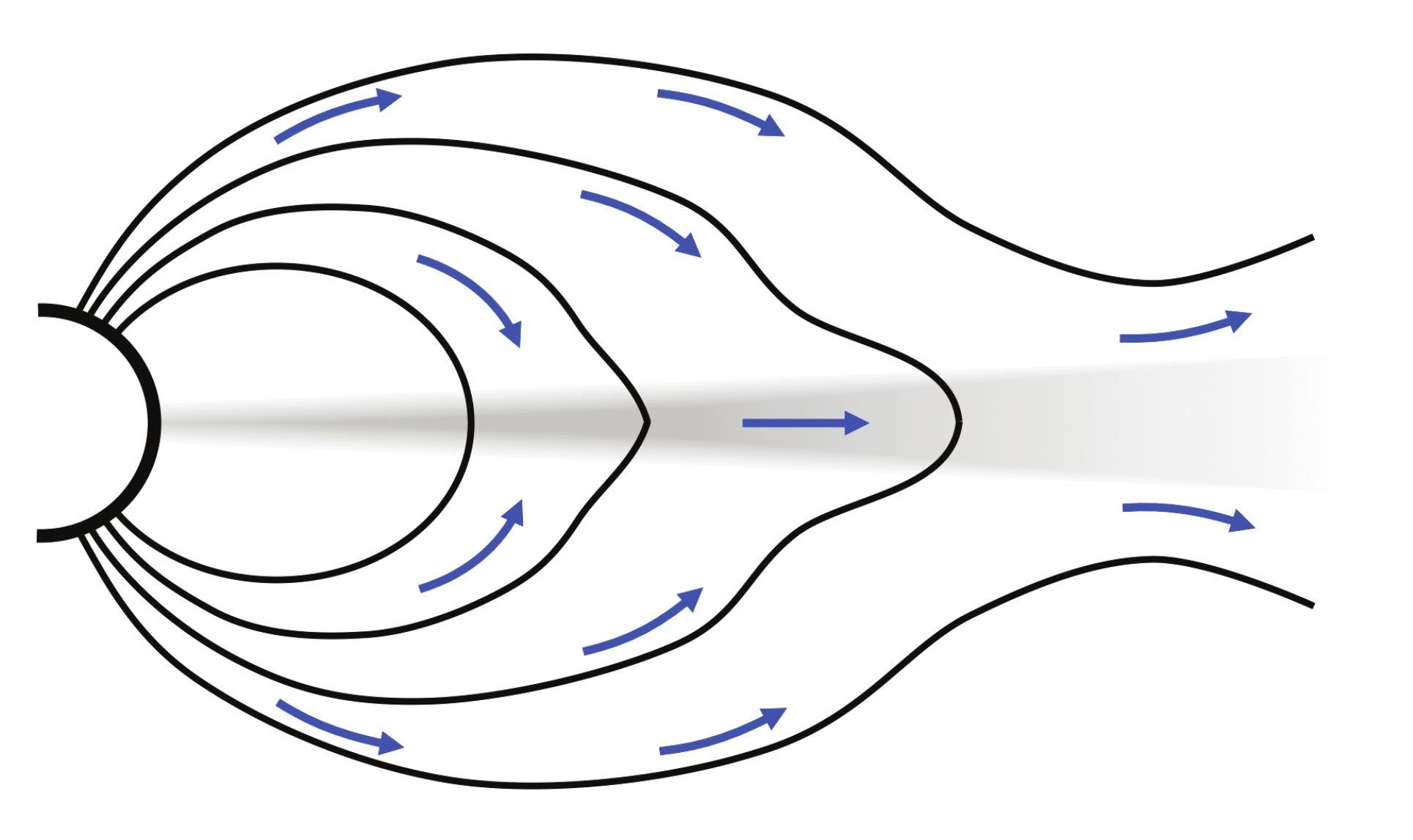}
\caption{
Illustration of centrifugal breakout (CBO) as a mode for emptying a stellar magnetosphere (gray shaded region) fed by a stellar wind mass upflow (blue arrows).
Adapted from \citet{Havnes84} and \citet{Owocki18}.
}
\label{fig:fig1}
\end{center}
\end{figure}

Since the development of this RRM model for CMs, a key question has been what loss processes eventually balance the steady feeding of the CM by the stellar wind.
For the simple case of field-aligned rotation, the appendices of 
TO05
presented an analytic analysis of the `centrifugal break out' (CBO) expected to occur \citep[e.g.,][]{Havnes84} 
when the CM reaches a {\em maximum confined mass}, above which the centrifugal force
overwhelms the magnetic tension that is confining disk material; see figure \ref{fig:fig1}.
A key result is that over a long term
this maximum confined mass
 depends on the magnetic field strength and Kepler radius, but is {\em independent} of the wind feeding rate.
Subsequent  two-dimensional (2D) magneto-hydrodynamical (MHD) simulations by \citet{udDoula06} and \citet{udDoula08} provided general support for the basic predictions of the semi-analytic RRM model and the associated analysis for CBO. 

The appendix of TO05 provided the basis for computing 
 this maximum confined mass and the associated
density distribution based on the CBO process.
However, the RRM model used in the body of the paper to derive empirical diagnostics simply assumed a density distribution set by the local mass feeding rate by the stellar wind, assuming a fixed but unspecified feeding time since the most recent emptying of the magnetosphere.
Our analysis here now derives a specific scaling relation for this 
 maximum confined mass
and density from CBO;
see equation (\ref{eq:sigr}), and section 2.2 for further elaboration on this comparison.

This CBO narrative has been challenged by \citet{Townsend13}, based on analysis of observations by the {\tt MOST} satellite of  the photometric variability in the prototypical RRM star $\sigma$~Ori~E.
They noted that the steady periodicity of these variations show no evidence of the large-scale disruption from CBO events seen in  2D-MHD simulations  by \citet{udDoula06} and \citet{udDoula08}; but subsequent 3D simulations \citep{udDoula13,Dalye-Yates19} show that such CBO eruptions are randomised over multiple azimuths around the star, and so upon spatial averaging would exhibit a much reduced overall variability.
However, \citet{Townsend13} also argue that the overall mass inferred from circumstellar absorption is substantially below (by a factor $\sim$50) that predicted from the CBO analysis.
The latter argument led \citet{Owocki18} to develop an alternative model for CM leakage, based on a steady, gradual diffusion and drift across turbulent magnetic field lines.

The present paper explores a new empirical diagnostic for discriminating between these competing 
scenarios for mass balance in CMs,
based on the recent analysis by \citet{Shultz20} of Balmer-$\alpha$ emission from the CMs around a broad sample of early B-type stars.
A particularly surprising result is given in the middle and right panels of their figure 3, which show that,
 above a threshold in luminosity or effective temperature,
the onset of detectable H$\alpha$ emission is {\em independent} of the stellar luminosity.
Since the radiatively driven stellar wind mass flux that feeds the CM should depend strongly on luminosity, this greatly disfavors a diffusive/drift leakage scenario, since that predicts an equilibrium CM mass that depends explicitly on the wind feeding rate \citep{Owocki18}.
Instead, this seems to favor the CBO paradigm for mass balance, for which the analysis in TO05 predicts an equilibrium CM mass that depends on the magnetic field strength and location of the Kepler radius, but is independent of wind feeding rate.

Building on this previous TO05 analysis, we show below (section 2.1) that this CBO model predicts a disk surface density that scales with the ratio of the magnetic energy density to stellar gravity at the Kepler radius, $\sigma_{\rm K} \sim B(R_{\rm K})^2/g(R_{\rm K})$, with then an associated optical thickness in H$\alpha$ that scales as $\tau_{\rm K} \sim \sigma_{\rm K}^2 \sim B_{\rm K}^4$.
This strong dependence of disk optical depth on the field strength at the Kepler radius helps explain the sudden onset of H$\alpha$ seen near a critical value $B(R_{\rm K}) \approx 100$\,G, as shown in the right panel of figure 3 from \citet{Shultz20}.
We next present (section 2.2) an extended analysis of the MHD simulations for CMs from \citet{udDoula08}, and use this to calibrate the analytic CBO scalings for
 maximum confined mass
 and its radial distribution above the Kepler radius.
With this calibration, we find (section 2.3) that the critical condition for optically thick disk emission is near the empirically inferred onset at $B_{\rm K} \approx 100$G.
For a power-law fit to this radial distribution of disk surface density, we then derive theoretical scalings for 
disk emission and the associated 
emission line profiles for both isothermal (section 3.1) and non-isothermal (section 3.2) models, and show that the former give an overall form that corresponds well with the empirical results from \citet{Shultz20}.
By integrating over the line profile we derive (section 3.3) the emission equivalent width and its ``curve of growth'' with increasing disk optical depth, and its strong dependence on the field strength at the Kepler radius.
We then (section 4) carry out explicit comparisons of the predictions of the  CBO model with the observational properties of the sample magnetic B-type stars analyzed by \citet{Shultz20}. This includes:  the onset of emission in early- to mid-B type stars  (section 4.1);
 the lack of emission in later B and A-type stars (section 4.2); 
 and the scaling of emission equivalent width with stellar and magnetic parameters (section 4.3).
After discussing (section 5)  the implications of this CBO model in the context of previous arguments against it (section 5.3 and 5.4), we conclude (section 6) with a summary of key results and an outline of directions for future work. 
Appendix A details our analyses of the strength of Balmer line opacity based on LTE vs. nebular recombination models.

\section{Centrifugal Breakout Scalings}

\subsection{Disk surface density and optical depth}

Following the appendices of TO05, we ground our analysis of CM mass distribution and its associated Balmer emission on the simple aligned-dipole case,
for which the angle between the rotational and magnetic axes $\beta = 0$.
In this case, the CM accumulation surface is a simple disk in the common equatorial plane for both the magnetic field and stellar rotation, with centrifugal support maintained for all radii $r$ at or above the Kepler co-rotation radius, $R_{\rm K}$ ($ \equiv  (G M/\Omega^2)^{1/3}$, for stellar mass $M$ and rotation frequency $\Omega$, with gravitation constant $G$). 

Equation (A4) of TO05 gives an expression for a characteristic surface density $\sigma_\ast$ for CBO\footnote{We retain the notation $\sigma_\ast$ from TO05. This characteristic surface density, and the associated disk optical thickness $\tau_\ast$ defined in eqn.\ (\ref{eq:taus}), are near, but not equal to, the associated MHD-calibrated values at the Kepler radius, $\sigma_{\rm K} = \sigma (R_{\rm K} )$ and $\tau_{\rm K  } = \tau_o (R_{\rm K} )$, as set by equations \ref{eq:sigr} and \ref{eq:taur}.}.
Applying the symbol definitions given there, 
this can readily be translated into an expression for surface density as a function of the field strength $B_{\rm K}$ and gravity $g_{\rm K}=GM/R_{\rm K}^2$ at the Kepler radius $R_{\rm K}$,
\beq
\boxed
{
	\sigma_\ast = \frac{B_{\rm K}^2}{4 \pi g_{\rm K}}
\, .
}
\label{eq:sigk}
\eeq
Multiplying both sides by the gravity $g_{\rm K}$, and noting that the magnetic term is related to the magnetic pressure $P_{\rm B} = B_{\rm K}^2/8 \pi$, we see that this has a similar scaling to that for 
hydrostatic equilibrium, wherein the pressure at any level is just given by gravity times the column mass of material above, $P  = \sigma g$.

The high radiation temperature of B-stars means that any circumstellar Hydrogen  in their CMs should be nearly fully ionised, with any observed Balmer-$\alpha$ emission arising from recombination cascade that includes transition from level 3 to 2.
Since the associated recombination rate depends on the product of the number densities of protons and electrons,  which scales with the square of the mass density  as $n_{\rm e} n_{\rm p} \sim \rho^2$, the associated emissivity likewise scales as $\eta \sim \rho^2$.
In terms of an associated absorption opacity $\kappa$ and absorptivity $\kappa \rho$, we can then define a source function $S \equiv \eta/(\kappa \rho)$, which for an LTE process like recombination is generally set by the temperature-dependent Planck function.
Assuming the disk temperature $T$  at any radius $r$ does not vary much with height $z$ above the disk,
the wavelength dependent specific intensity (a.k.a. surface brightness) $I_\lambda$ from the disk\footnote{To compare with observational analyses that subtract the background absorption profile from the underlying star, we concentrate here on only the intrinsic emission from the disk, assuming that this dominates any scattering of stellar radiation.} is then given by the formal solution for radiative transfer,
\beq
I _\lambda (r,\mu) = S (r)  \left [ 1 - e^{-\tau_\lambda (r)/\mu} \right ]
\, ,
\label{eq:formsoln}
\eeq
where $\tau_\lambda (r) $ is the  line optical depth at wavelength $\lambda$ through the disk normal at radius $r$, while $\mu$ is the projection cosine of the line-of-sight to the disk normal.

Within the TO05 RRM model, the volume density at any given radius $r$ has a gaussian stratification from its midplane value $\rho_{\rm m} (r)$,
\beq
\rho (z,r) = \rho_{\rm m} (r) e^{-(z/h)^2} 
\, ,
\label{eq:rhozr}
\eeq
where $h$ is a characteristic scale height. 
Writing the line-center opacity as $\kappa_{\rm o} = C_{\rm o} \rho$, where $C_{\rm o}$ is a coefficient derived from the specific radiative transfer model (see Appendix A),
the optical depth is given by integration of $\kappa_{\rm o} \rho = C_{\rm o} \rho^2$ over the full height range ($-\infty < z < \infty $) through the disk.
Using the fact that $\sigma = \rho_{\rm m} h \sqrt{\pi}$, we find the associated line-center optical depth near the Kepler radius scales as
\beq
	\tau_\ast 
	= \frac{C_{\rm o} \sigma_\ast^2 }{\sqrt{2} \,  \pi h_{\rm K}} 
	= \frac{C_{\rm o} }{16 \sqrt{2} \pi^3 h_{\rm K}} \, \frac{  B_{\rm K}^4}{g_{\rm K}^2}  
	= \frac{C_{\rm o} }{32 \pi^3 c_{\rm s}} \, \frac{ B_{\rm K}^4 R_{\rm K}^{5/2}}{(GM)^{3/2}} 
\, .
\label{eq:taus}
\eeq
Here the second equality uses  equation (\ref{eq:sigk}) and the fact that the disk scale height at the Kepler radius depends on the sound speed $c_{\rm s}$ and 
stellar rotation frequency $\Omega \equiv  2 \pi /P$,
\beq
	h_{\rm K} = \frac{\sqrt{2} c_{\rm s}}{\Omega} 
\, ,
\label{eq:hK}
\eeq
and the final equality uses the fact that $\Omega = \sqrt{g_{\rm K}/R_{\rm K}}$.

A  key result from equation (\ref{eq:taus})  is that this disk optical depth near the Kepler radius depends very sensitively on the field strength length there, scaling as $\tau_\ast \sim B_{\rm K}^4$.
Because in the RRM model the disk surface density has its maximum value near the Kepler radius, its optical depth likewise is greatest there, and so first crosses from optically thin to thick when $\tau_\ast \approx 1$.
For an optically thin disk with $\tau_\ast < 1$, expansion of the formal solution (\ref{eq:formsoln}) shows that the  normal ($\mu=1$) surface brightness scales as $I_1 (R_{\rm K}) \approx S(R_{\rm K}) \tau_\ast $, which is a factor $\tau_\ast \ll 1$ less than the saturated brightness $I_1 (R_{\rm K}) = S(R_{\rm K})$ in the optically thick limit $\tau_\ast \gg 1$.
Since $\tau_\ast \sim B_{\rm K}^4$, this helps explain the sudden onset of detectable emission at a critical value of $B_{\rm K}$, as shown in the rightmost panel of figure 3 from \citet{Shultz20}.

\begin{figure}
\includegraphics[scale=.4]{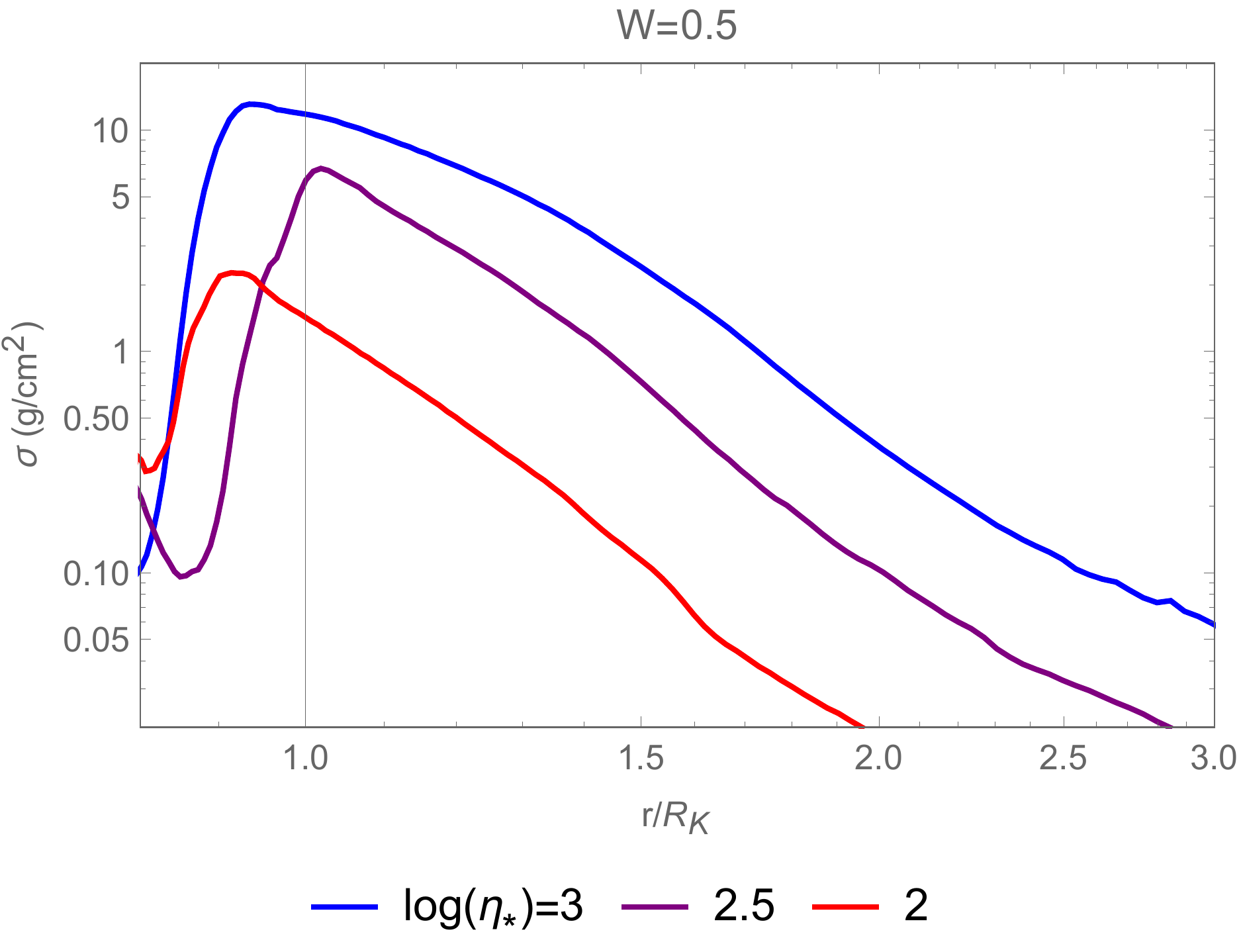}
\includegraphics[scale=.4]{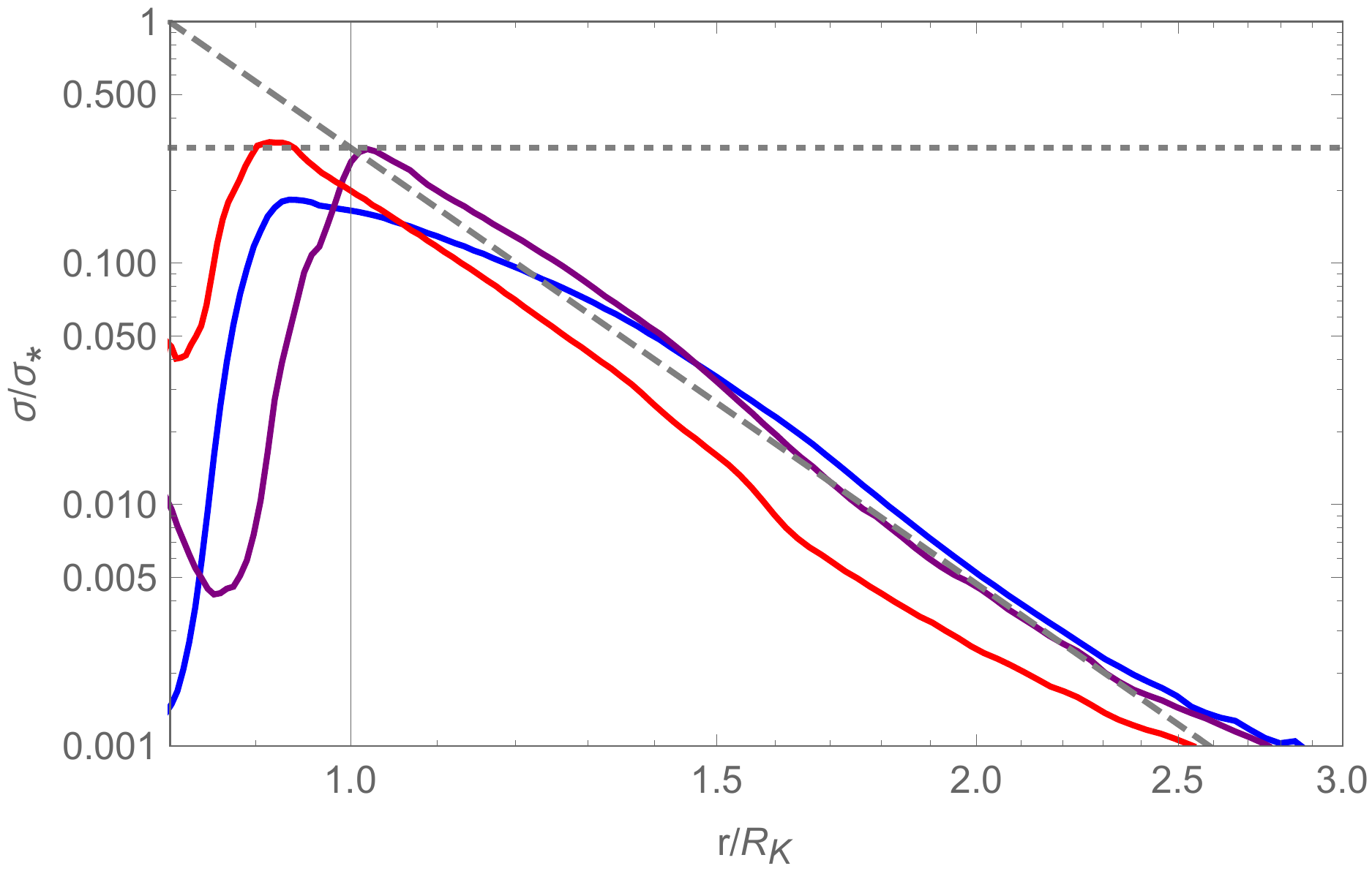}
\caption{{\em Upper panel:} 
Disk surface density averaged over the last 1.5\,Msec for the 2D-MHD simulations from \citet{udDoula08}, plotted vs. radius over Kepler radius, $r/R_{\rm K}$, for a model series with ratio of equatorial rotation speed to surface orbital speed, $W \equiv V_{\rm rot}/V_{orb} = 0.5$, giving a Kepler corotation  radius to stellar radius of $R_{\rm K}/R_\ast = W^{-2/3}=1.6$.
The legend marks the colors for the various assumed values of the magnetic confinement parameter $\eta_\ast$.
{\em Lower panel:}
For these same
3 cases with strongest field,
the radial variation of MHD surface density normalised by analytic Kepler radius values in equation (\ref{eq:sigk}).
The thin vertical line shows that
the intermediate confinement model (purple curve) 
 has a peak very near the Kepler radius, but the horizontal dotted line shows that the peak value is a factor 0.3 lower than analytic predication for $\sigma_\ast$. 
 The diagonal dashed line shows the radial decline from the Kepler radius closely follows a power law, $(r/R_{\rm K})^{-q}$, with $q \approx 6$.
}
\label{fig:fig2}
\end{figure}

\subsection{Calibration by MHD simulations}

Following the TO05 development of the RRM model for CMs, \citet{udDoula08} carried out full 2D numerical MHD simulations of magnetically confined stellar winds for the axisymmetric case of aligned-dipole stellar rotation. 
For cases with sufficiently strong field confinement and rapid rotation --  {\em i.e.} with Kepler radius well within the Alfven radius, $R_{\rm K} \ll R_{\rm A}$ --, results showed the overall time-averaged, equilibrium mass in the resulting CM agrees quite well with the predicted analytic scalings  for CBO given by TO05 equation (A11).

To facilitate computation here of Balmer emission from such CMs, we have now derived associated results for the time-averaged surface density from these same 2D MHD simluations.
Figure 7 of  \citet{udDoula08}  plots the disk mass distribution $dm/dr$ vs. radius and time, for a mosaic of models with various rotation parameters  $W \equiv V_{\rm rot}/V_{orb} $ and magnetic confinement parameters, $\eta_\ast \equiv B_\ast^2 R_\ast^2/{\dot M} v_\infty$, with the solid and dashed horizontal lines showing the associated Alfven radii $R_{\rm A} \sim \eta_\ast^{1/4}$ and Kepler radii $R_{\rm K} \sim W^{-2/3}$.

For the case with moderately rapid rotation, $W =1/2$ ($R_{\rm K} = 1.59 R_\ast$), the top panel of figure \ref{fig:fig2} here now shows the associated time-averaged (over the final 1500\,ks of the simulations) surface density $\sigma = (dm/dr)/(2 \pi r)$ 
for the strong confinement cases $\log \eta_\ast=$2, 2.5 and 3, plotted on a log-log scale vs.\ the radius scaled by the Kepler radius, $r/R_{\rm K}$.
The central legend shows the line style for each value of $\log \eta_\ast$.
Note that for all these strong confinement cases, 
the surface density peaks near the Kepler radius, then shows nearly a linear decline outward on this log-log scale, indicating a power law.

The lower panel of figure \ref{fig:fig2} rescales these surface densities by the analytic breakout value from equation (\ref{eq:sigk}), $\sigma/\sigma_\ast$.
The horizontal dotted line at a value of 0.3 shows that the peak density at the Kepler radius is actually reduced by about 30\%  from the analytic value given in equation (\ref{eq:sigk}).
The slanting dashed line shows a power-law decline from this Kepler value, $\sigma/\sigma_\ast = 0.3 (r/R_{\rm K})^{-q}$,  with slope\footnote{
This is significantly steeper than the $q=3$ index generally assumed for the RRM model, based on an assumption of a fixed time for wind feeding at a rate ${\dot \sigma} \sim B(r) \sim 1/r^3$; see section 5.2.} $q=6$.
The fit to the intermediate confinement case $\log \eta_\ast = 2.5$ is quite good. For $\log \eta_\ast =$2 and 3, the peaks occur somewhat below $r=R_{\rm K}$, but even for these cases, the radial declines closely parallel the dashed line.

To compare this further with the TO05 analytic scalings for CBO,  note that applying their equation (A4) into their (A3) gives for the variation of breakout density with scaled radius $\xi \equiv r/R_{\rm K}$,
\beq
\sigma_{\rm b} (\xi) = \frac{\sqrt{\pi} \sigma_\ast}{\xi^4 \left ( \xi^3 - 1 \right )}
\, ,
\label{eq:a3}
\eeq
which at radii $\xi \gg 1$ far above the Kepler radius approaches power-law with an index $q=7$,  only slightly steeper than the above empirical fit with $q=6$.
But because the net gravito-centrifugal force is by definition zero at the Kepler radius, this analytic CBO scaling implies that the breakout density should formally diverge at $r=R_{\rm K}$ ($\xi=1$).
In practice, figure 7  of  \citet{udDoula08} shows that in the MHD simulations any breakouts that occur above the Kepler radius lead to disturbances in the overall magnetosphere that
induces {\em infall} back to the star from the region around the Kepler radius.
The net result is a finite surface density there that is actually somewhat {\em below} (by about a factor 0.3) the scaling form (\ref{eq:sigk}), with a radial drop off that is slightly less steep, i.e., with power-index $q=6$ instead of $q=7$.

For the emission model computations in this paper, we thus adopt an MHD-calibrated scaling for the surface density given by
\beq
\boxed{
\sigma (r) = 0.3  \, \sigma_\ast \left ( \frac{r}{R_{\rm K}} \right )^{-6} 
=  0.3 \frac{B_{\rm K}^2}{4 \pi g_{\rm K}} \, \left ( \frac{r}{R_{\rm K}} \right )^{-6} 
\, .
}
\label{eq:sigr}
\eeq
This contrasts with the RRM scaling invoked by TO05, which assumed the entire magnetosphere was last emptied at some unspecified fixed time  $t_{\rm e}$ in the past.
That gives a radial variation of surface density that is proportional to the local stellar-wind feeding rate, which for  a flow along a closed dipole flux loop scales as 
$ \sigma (r) \sim \dot{\sigma}(r)  t_{\rm e} \sim B(r) \sim r^{-3}$.
By comparison, the CBO model here now gives a steeper radial decline, $\sigma (r) \sim B^2 (r) \sim r^{-6}$, with moreover a given overall density, instead of invoking an unspecified filling time that leaves the overall density likewise unspecified.

With this full CBO scaling (\ref{eq:sigr}), we can generalise equation (\ref{eq:taus}) to obtain the radial variation of line-center optical depth,
\beq
\tau_{\rm o} (r) = \frac{C_{\rm o} \sigma^2 (r)}{\sqrt{2} \pi h(r)} 
= 0.09 \tau_\ast \, \left ( \frac{r}{R_{\rm K}} \right )^{-12}  \, \frac{h_{\rm K}}{h(r)}
\, ,
\label{eq:taur}
\eeq
where the scale height variation $h_{\rm K}/h(r) = \sqrt{3 - 2 (R_{\rm K}/r)^3}$ ranges  from unity at $R_{\rm K}$ to a factor $\sqrt{3}$ at large radii  \citep[][equation 4]{Owocki18}.
Apart from this modest radial increase, the optical depth drops very steeply with radius, as $\tau_{\rm o} \sim r^{-12}$,
leading to quite sharp outer edges to disk emission (see figure \ref{fig:fig4} below).

\subsection{Critical field evaluation}

With this MHD calibration for disk density and thus optical depth, let us next determine the critical field strength $B_{\rm K1}$  for making the disk become optically thick.
For this, we first apply the analysis in the Appendix to derive scalings for the coefficient $C_{\rm o}$ in equation (\ref{eq:taus}).
From (\ref{eq:taur}), we have $\tau_{\rm K} \equiv \tau_{\rm o} (R_{\rm K}) = 0.09 \tau_\ast \sim B_{\rm K}^4$, 
so the critical field condition $\tau_{\rm K} =1$ solves to
\beqa
B_{\rm K1} 
	&=&  \left (  \frac{16 \sqrt{2} \, \pi^3 h_{\rm K}}{0.09 \, C_{\rm o}} \right )^{1/4}  \sqrt{g_{\rm K}} 
\label{eq:BK1a}
\\
	&\approx&  \left (70 {\rm G} \, T_{\rm 20kK} + \Delta B \right ) \, \sqrt{g_{K4}} ~ P_{\rm day}^{1/4}
\,  .
\label{eq:BK1b}
\eeqa
The latter equality gives an analytic fit form, with $T_{\rm 20 kK} \equiv T/20 kK$,
$g_{K4} \equiv g_{\rm K}/(10^4 \, {\rm cm/s^2})$, 
$P_{\rm day}$ the rotation period in days, and 
$\Delta B$\ an optional offset for using the nebular vs.\ LTE model for H$\alpha$ opacity discussed in the Appendix.

For $P_{\rm day} = g_{K4} = 1$,
figure \ref{fig:fig3} compares the variation of $B_{\rm K1}$ vs. disk temperature $T$ for 
the nebular (red) vs. LTE (blue) models for the Balmer-line opacity;
note they both have a roughly linear increase with temperature, with the nebular value showing a nearly fixed offset $\Delta B_{\rm neb} \approx +90$\,G.

Indeed, the average between the LTE and nebular models (shown by the dashed line in figure \ref{fig:fig3}) can be approximated simply by setting $\Delta B =  \Delta B_{\rm neb}/2 = 45$\,G in equation (\ref{eq:BK1b}). This average gives typical critical fields $B_{\rm K1} \approx 100 $\,G,  in remarkably good agreement with the empirically inferred value for onset for H$\alpha$ emission, as shown in the right panel of figure 3 of from \citet{Shultz20}.
Overall then, choosing $\Delta B =$0, 45, or 90~G represents respectively  the LTE, average, or nebular models for opacity.

\begin{figure}
\begin{center}
\includegraphics[scale=0.45]{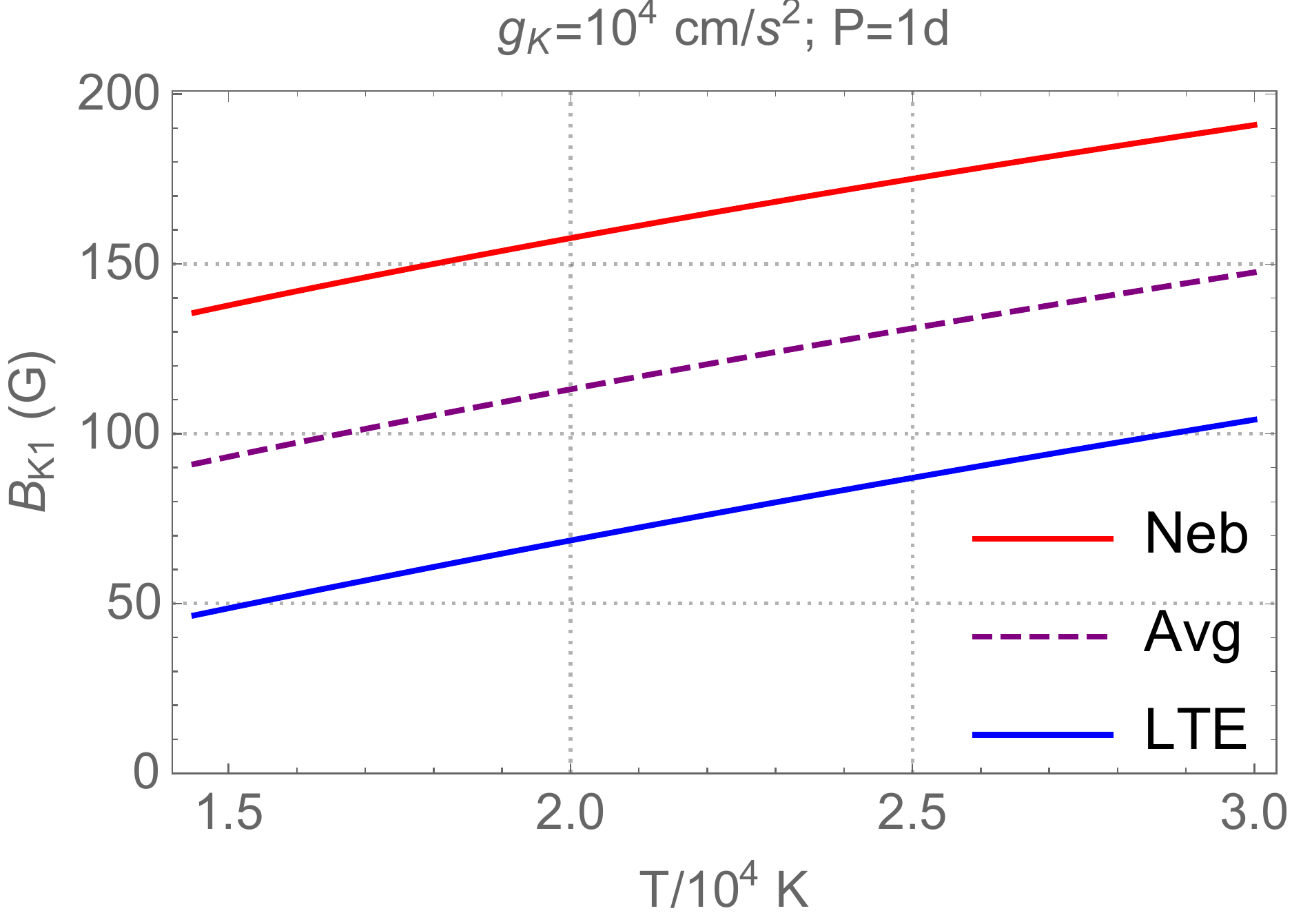}
\caption{For a star with rotation period $P=1$\,day and Kepler radius gravity $g_{\rm K} = 10^4$\,cm\,s$^{-2}$, the critical field $B_{\rm K1}$ vs.\ disk temperature $T$ for both the nebular (red curve) and LTE (blue curve) models developed in Appendix A for H$\alpha$ formation. The purple dashed curve shows the average between the two models.
Equation (\ref{eq:BK1b}) gives a simple analytic scaling formula for $B_{\rm K1}$.
}
\label{fig:fig3}
\end{center}
\end{figure}

\section{Disk Emission}

\subsection{Line profiles for isothermal disk}

Let us now derive the emission line profiles associated with this analytic CBO model, through application of the optical depth $\tau_{\rm o} (r)$ from equation (\ref{eq:taur}) into the formal solution (\ref{eq:formsoln}). 
For this let us write the local wavelength dependence in terms of a line-profile function $\phi_\lambda = \tau_\lambda/\tau_{\rm o}$, with a small thermal Doppler width $\Delta \lambda_{\rm D} 
= \lambda_{\rm o} v_{\rm th}/c
$ about a line-center wavelength $\lambda_{\rm o}$.
For a simple box profile with $\phi_\lambda = 1$ if $|\lambda - \lambda_{\rm o} | < \Delta \lambda_{\rm D}/2$, and zero otherwise, the intensity along a direction cosine $\mu$ to the disk normal at radius $r$ is likewise
\beq
I_\lambda (r,\mu) = S(r) \left [ 1 - e^{-\tau_{\rm o} (r)/\mu} \right ]   \equiv I_{\rm o} ~ {\rm if} ~  | \lambda - \lambda_{\rm o} | < \Delta \lambda_{\rm D}/2
\, ,
\eeq
and zero otherwise.
The wavelength-integrated intensity is thus ${\bar I }= I_{\rm o} \Delta \lambda_{\rm D}$.

\begin{figure}
\includegraphics[scale=.275]{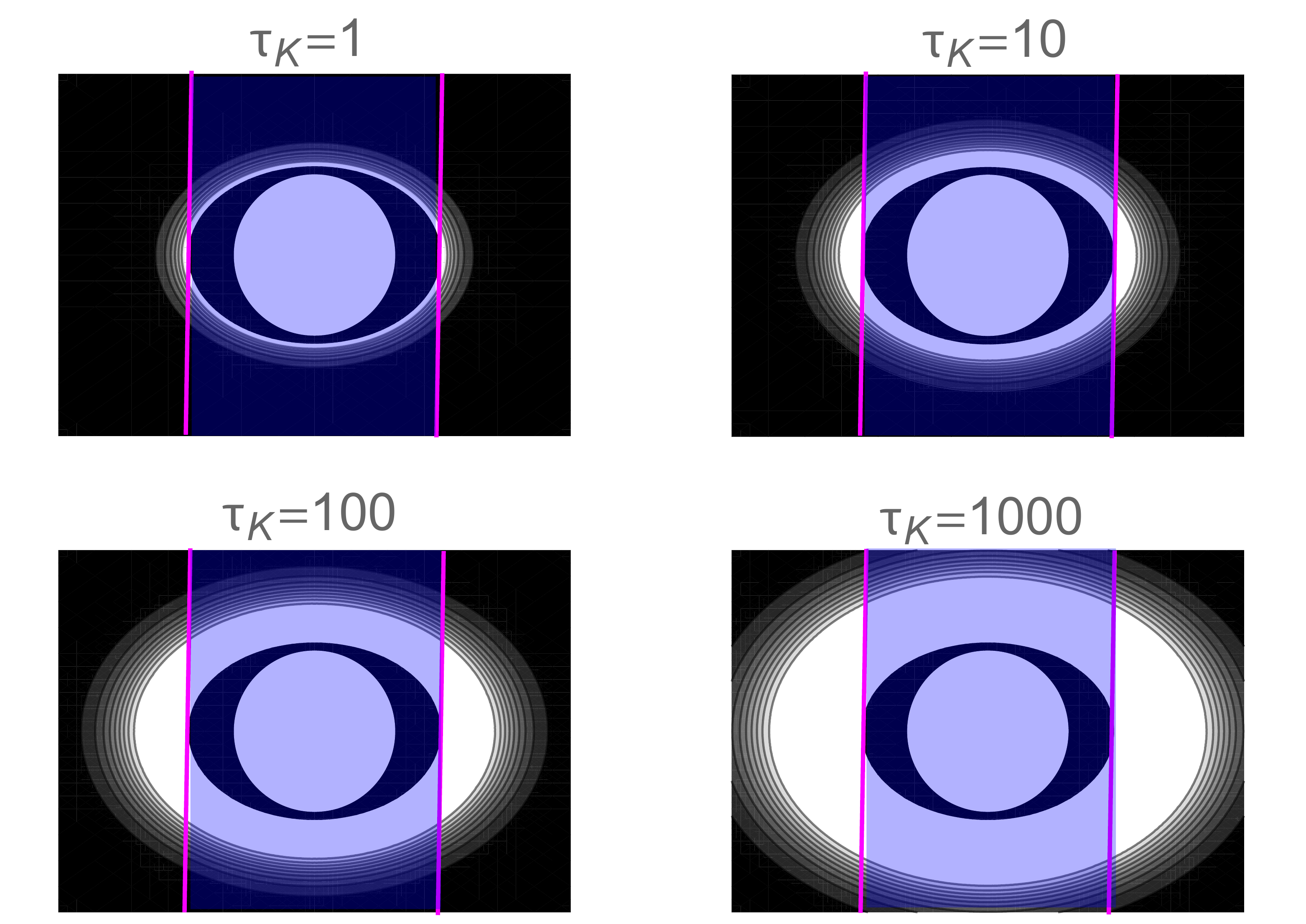}
\caption{Line-center surface brightness $I_{\rm o}$ for {\em isothermal} CM disks viewed from an inclination $i=45^o$, for rotation fraction $W=1/2$ and Kepler radius optical depths $\tau_{\rm K}=$1, 10, 100, and 1000.
For simplicity, the disk and central star are assumed to have the same temperature, and thus the same source function and saturated surface brightness $I_{\rm o}=S$.
The vertical magenta lines denote offsets by one Kepler radius to each side of  the central projected rotation axis.
To emphasise the purely disk emission outside the Kepler radius, both stellar and disk emission are dimmed inside the vertical lines at $\pm R_{\rm K}$.
}
\label{fig:fig4}
\end{figure}

\begin{figure}
\begin{center}
\includegraphics[scale=.45]{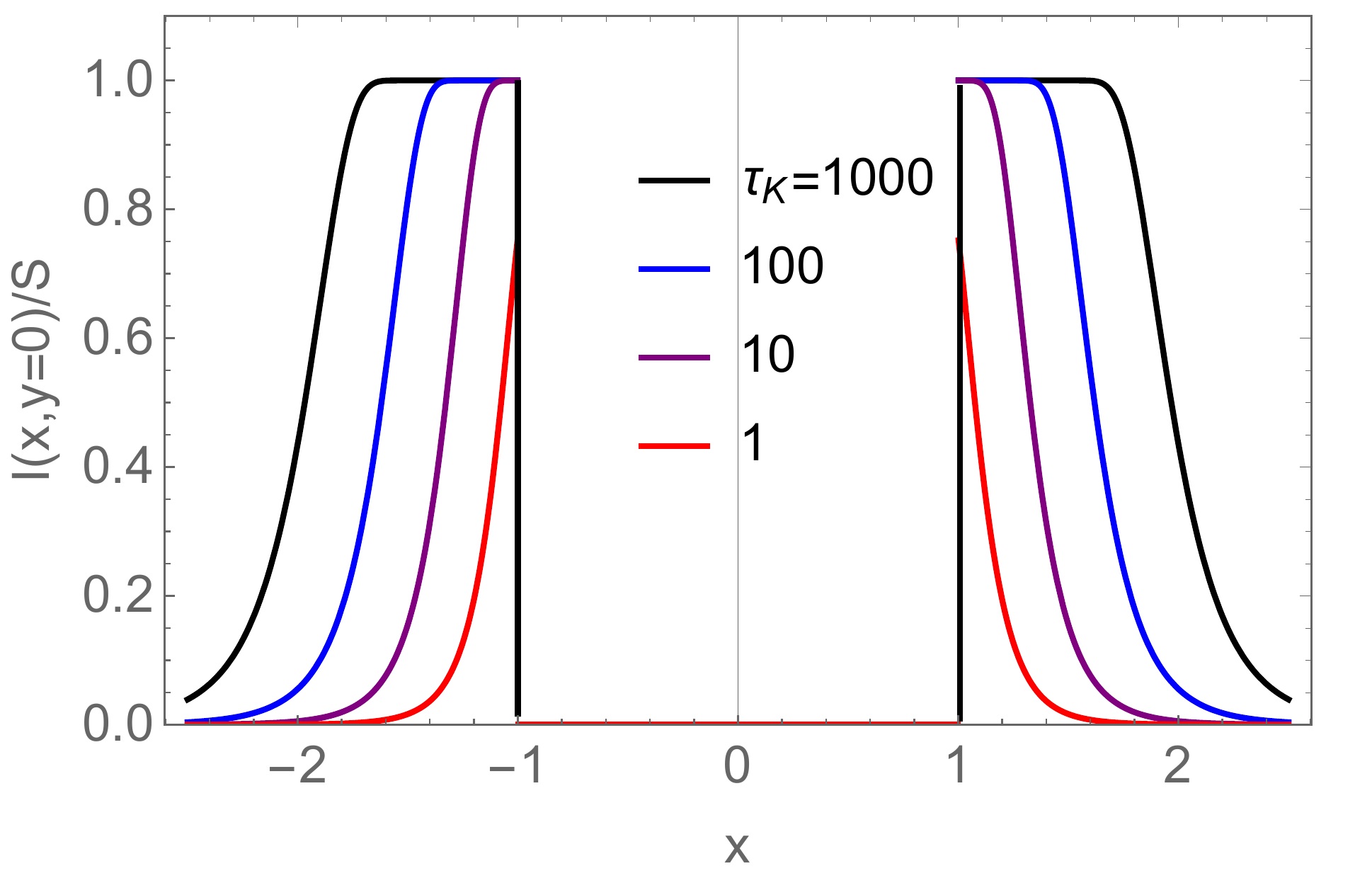}
\includegraphics[scale=.45]{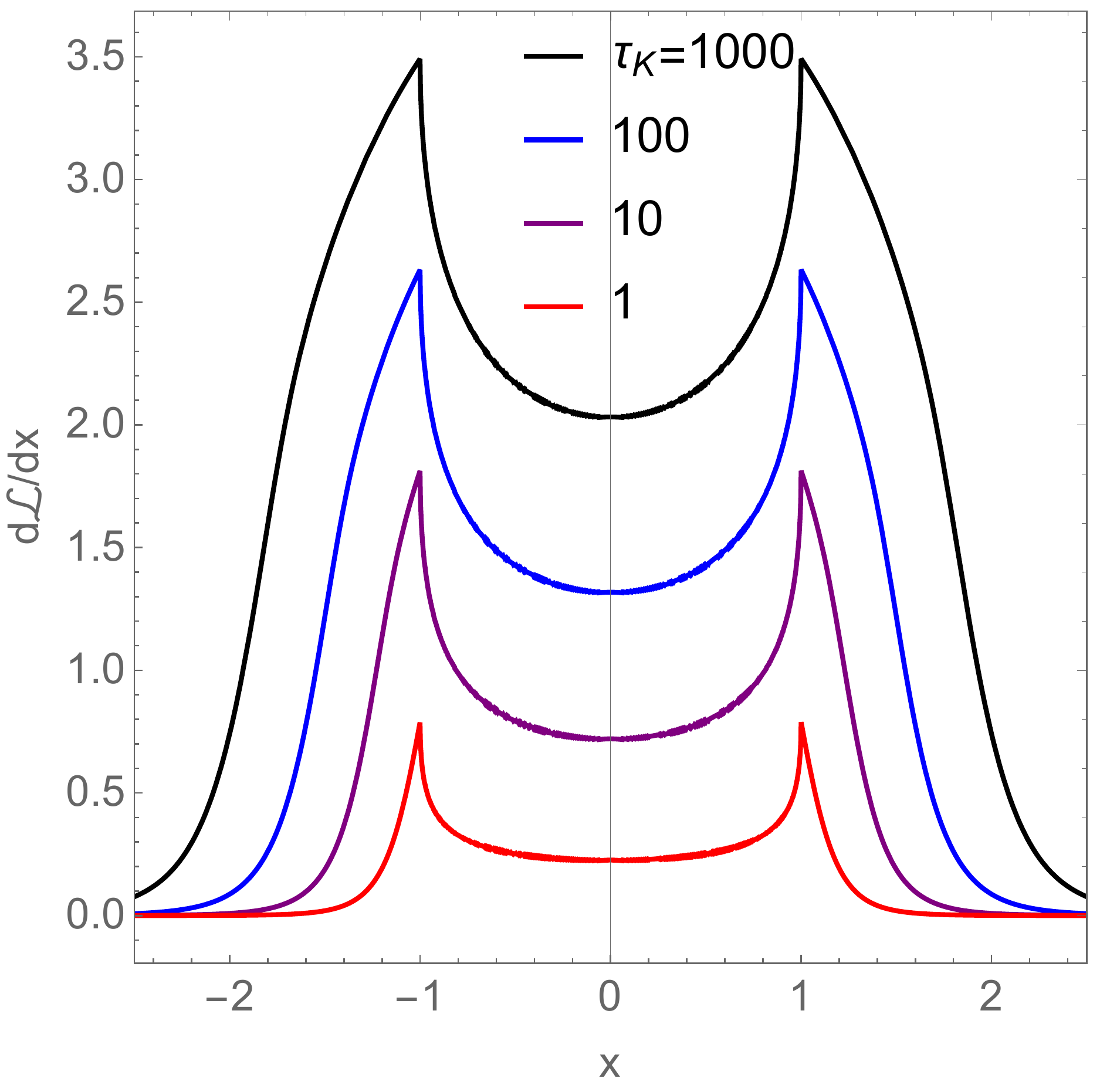}
\caption{
{\em Top:} Horizontal variation of disk surface brightness through the mid-plane, $I(x,y=0)$, plotted for the labeled values of $\tau_{\rm K}$.
{\em Bottom:} Associated emission line profile, as computed from the $y$-integration in equation (\ref{eq:Lx}), 
with $x$ now identified as representing the observed Doppler shift in units of the shift from the projected co-rotation at the Kepler radius.
}
\label{fig:fig5}
\end{center}
\end{figure}

As a first example, let us consider the simple case of an {\em isothermal} disk with spatially constant source function $S$, around a star with rotation fraction $W=1/2$ (giving $R_{\rm K}/R_\ast = W^{-2/3} = 1.6$), viewed from an intermediate inclination $i=45^o$ (with thus $\mu \equiv \cos i = 1/\sqrt{2}$).
The four panels in figure \ref{fig:fig4} illustrate the surface brightness $I _{\rm o} (x,y)$ projected onto the $(x,y)$ plane of the sky,
 for the four labeled values for the disk optical thickness at the Kepler radius, $\tau_{\rm K} \equiv \tau_{\rm o}(R_{\rm K}) = 0.09 \tau_\ast$.
The illustrations also include the central star, which for simplicity is assumed here to have the same temperature as the disk, with thus equal source function $S$ and so equal optically thick surface brightness $I_{\rm o}=S$ (ignoring limb darkening). 
The vertical ($y$) magenta lines denote offsets by one Kepler radius to each side of  the central projected rotation axis.
To emphasise the purely disk emission outside this radius, we have dimmed the stellar and disk emission inside these lines at $ \pm R_{\rm K}$, i.e., at $|x|<1$.

With $x$ and $y$ thus scaled by the Kepler radius, the top panel of figure \ref{fig:fig5} plots the horizontal ($x$) variation of surface brightness along the mid-plane, 
$I_{\rm o} (x,y=0,\mu)$.
The bottom panel compares a 
{\em vertically integrated} emission distribution in $x$, 
 \beq
\frac{d\mathcal{L}}{dx}
\equiv 
\frac{1}{\mu S_{\rm K} \Delta \lambda_{\rm DK}}
\int_{-\infty}^\infty {\bar I} (x, y,\mu) \, dy
\, .
\label{eq:Lx}
\eeq
where 
the factor $1/\mu $ corrects for the projected foreshortening of  differential vertical element $dy$, and the normalisation by the source function and Doppler width at the Kepler radius makes this fully dimensionless.  

For the assumed rigid-body rotation of this disk, the projected Doppler shift from co-rotation scales directly with the horizontal displacement $x$.
As such, this plot of $d\mathcal{L}/dx$ vs. $x$ can be equivalently interpreted as a {\em disk emission line profile}, with $x$ now representing the frequency displacement from line center in units of the Doppler shift associated with the projected rotation velocity at the Kepler radius, $\Omega R_{\rm K} \sin i$, which itself is just a known factor $R_{\rm K}/R_\ast$ higher than the projected stellar rotation velocity $V_{\rm rot} \sin i$.

Note that for disks that are strongly optically thick near the Kepler radius ($\tau_{\rm K} \ge 10$), the profiles in the inner wing region $|x| \gtrsim 1$ show a {\em concave down} shape, similar to the observed profile form from \citet{Shultz20} (see their figure 10, reproduced in the lower panel of figure \ref{fig:fig7} here).
 Such downward concavity follows from the fact that $y$-segments across the bright disk have the greatest length when nearly tangent to the Kepler radius.
 The near constancy of $I_{\rm o}=S$ over this optically thick segment reflects the assumed constancy  of the disk source function $S$, which for LTE emission in the Rayleigh-Jeans tail of the Planck function is directly proportional to the assumed-constant disk-temperature $T_d$.
 
 \subsection{Effect of a power-law radial decline in temperature}
 
 More generally, we can consider a temperature and resulting source function that declines as a power-law of index $s$ from its value at the Kepler radius,
 \beq
 S(r) = S_{\rm K} \left ( \frac{r}{R_{\rm K}} \right )^{-s}
 \, ,
 \label{eq:Sr}
 \eeq
 which can be similarly applied into the formal solution (\ref{eq:formsoln}) for the surface brightness $I (x,y)$.
 Since H$\alpha$ lies in the Rayleigh-Jeans tail for B-star temperatures, we have $S  \sim T$. 
Following the nebular model for opacity, $C_{\rm o} \sim T^{-1.5}$, we assume the disk optical depth scales as $\tau \sim C_{\rm o}/\sqrt{T}  \sim T^{-2} $.
For  surface density $\sigma  \sim  r^{-6}$, a non-zero $s$ gives the optical depth a somewhat less steep radial decline, $\tau \sim r^{2s-12}$.

\begin{figure}
\begin{center}
\includegraphics[scale=.145]{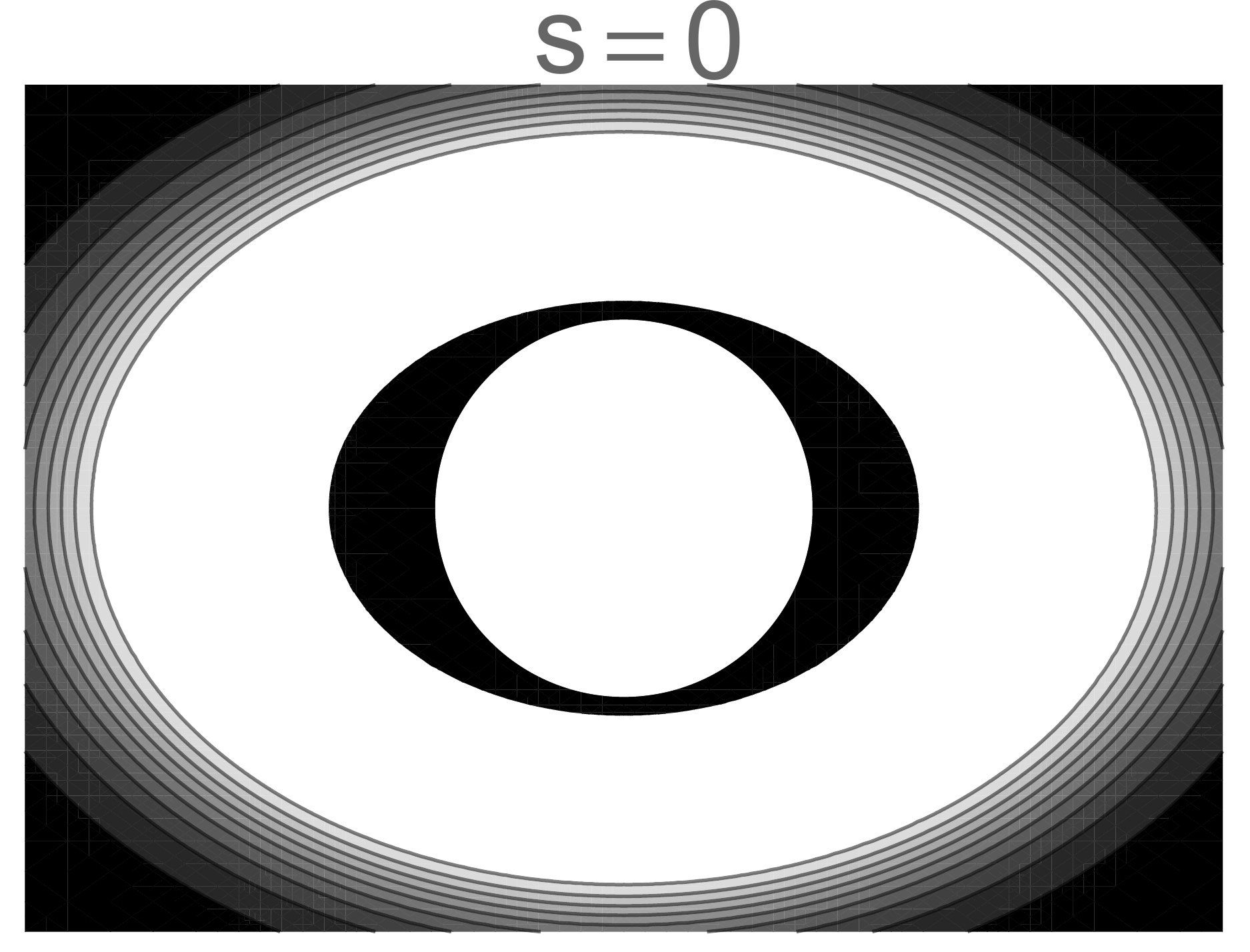}
\includegraphics[scale=.145]{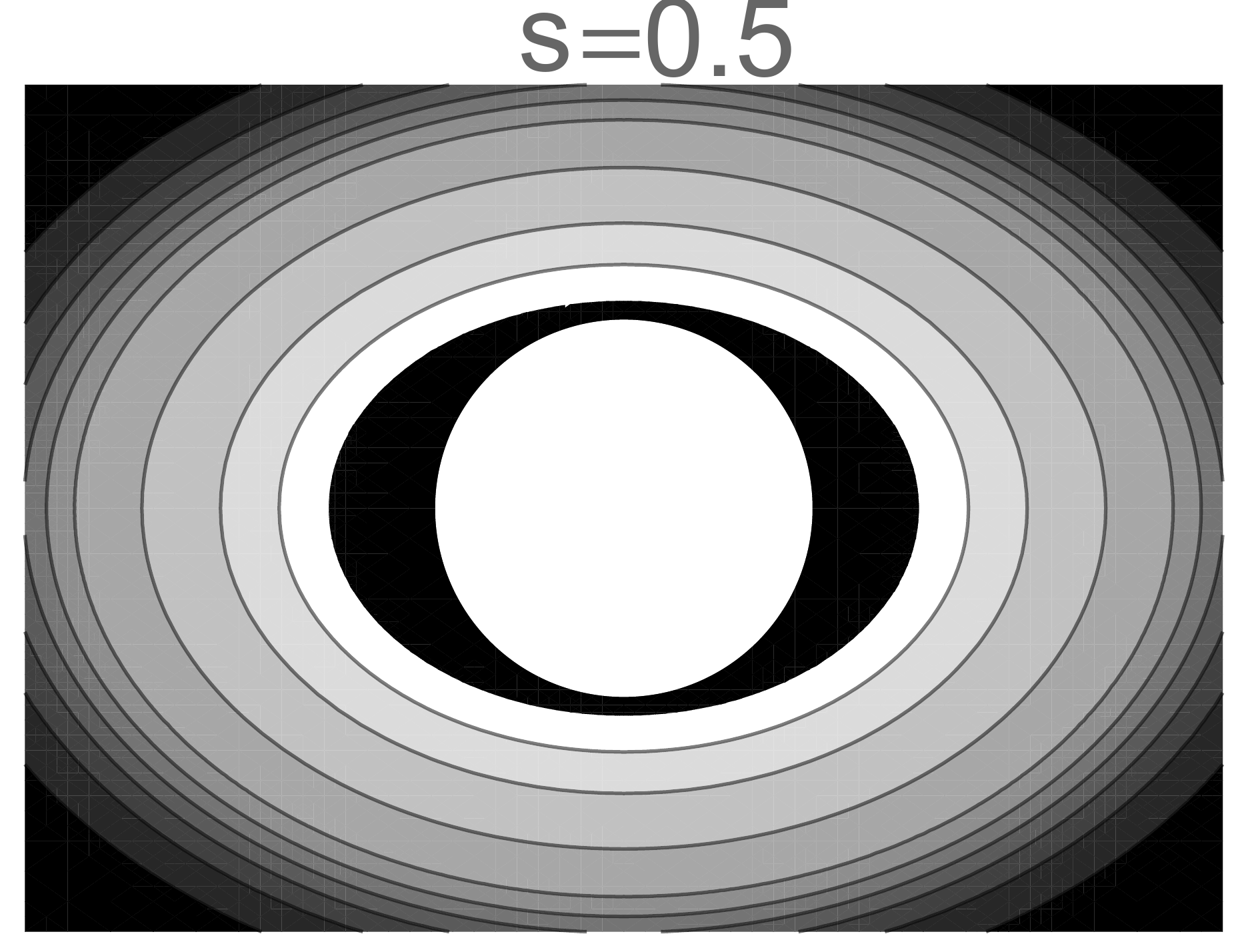}
\includegraphics[scale=.145]{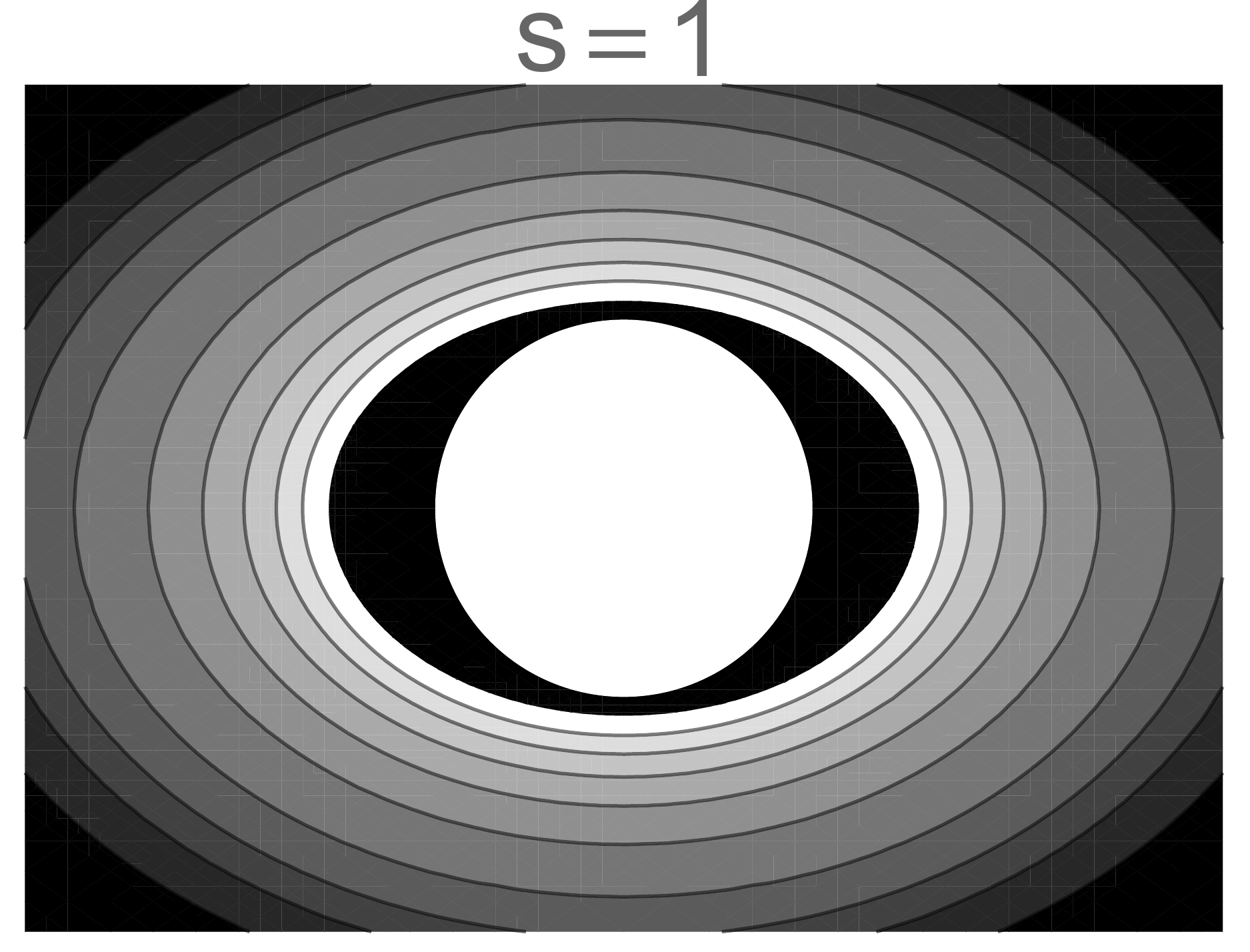}
\end{center}
\includegraphics[scale=.4]{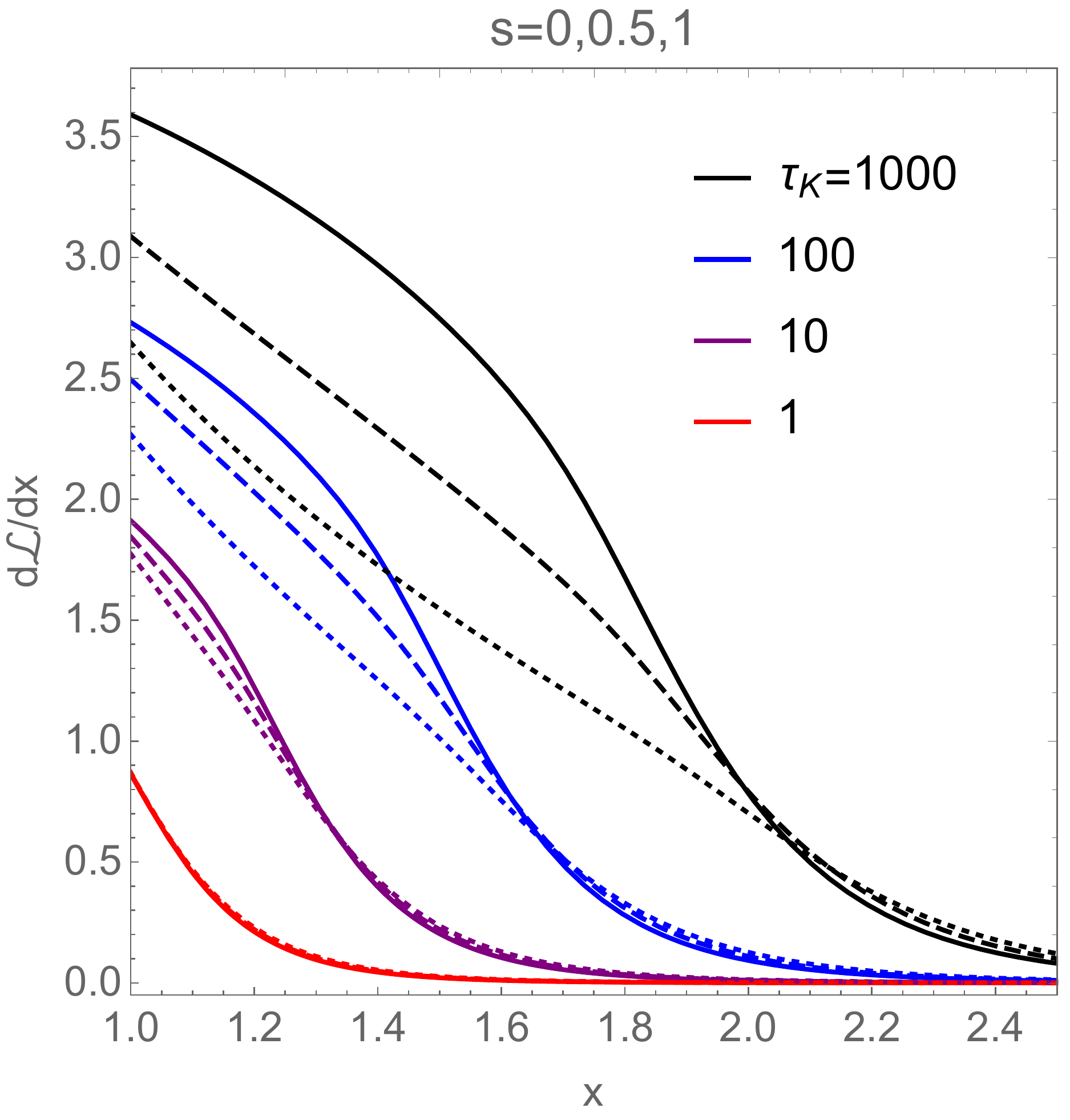}
\caption{{\em Top:} Same as figure \ref{fig:fig4}, but now comparing the isothermal case for largest optical depth $\tau_{\rm K}=1000$ with results for disks with source function declining as a power-law
$S(r)  = S_{\rm K} (r/R_{\rm K})^{-s}$, with $s=0$, 
0.5, and 1
 (left, middle, and right panels),
 assuming the disk optical depth scales as $\tau \sim C_{\rm o}/\sqrt{T} \sim T^{-2}$, 
as appropriate for the nebular model for opacity factor $C_{\rm o}$ derived in Appendix A.
The contours show variations in surface brightness $I(x,y)$ from unity downward in steps of $0.1$.
{\em Bottom:} Associated wing emission line profiles for cases with $s=0$, 0.5, 
and 1 (solid, dashed, and dotted) and $\tau_{\rm K}=1000$, 100, 10, and 1 (marked by legend colors).
}
\label{fig:fig6}
\end{figure}

\begin{figure}
\begin{center}
\includegraphics[scale=.35]{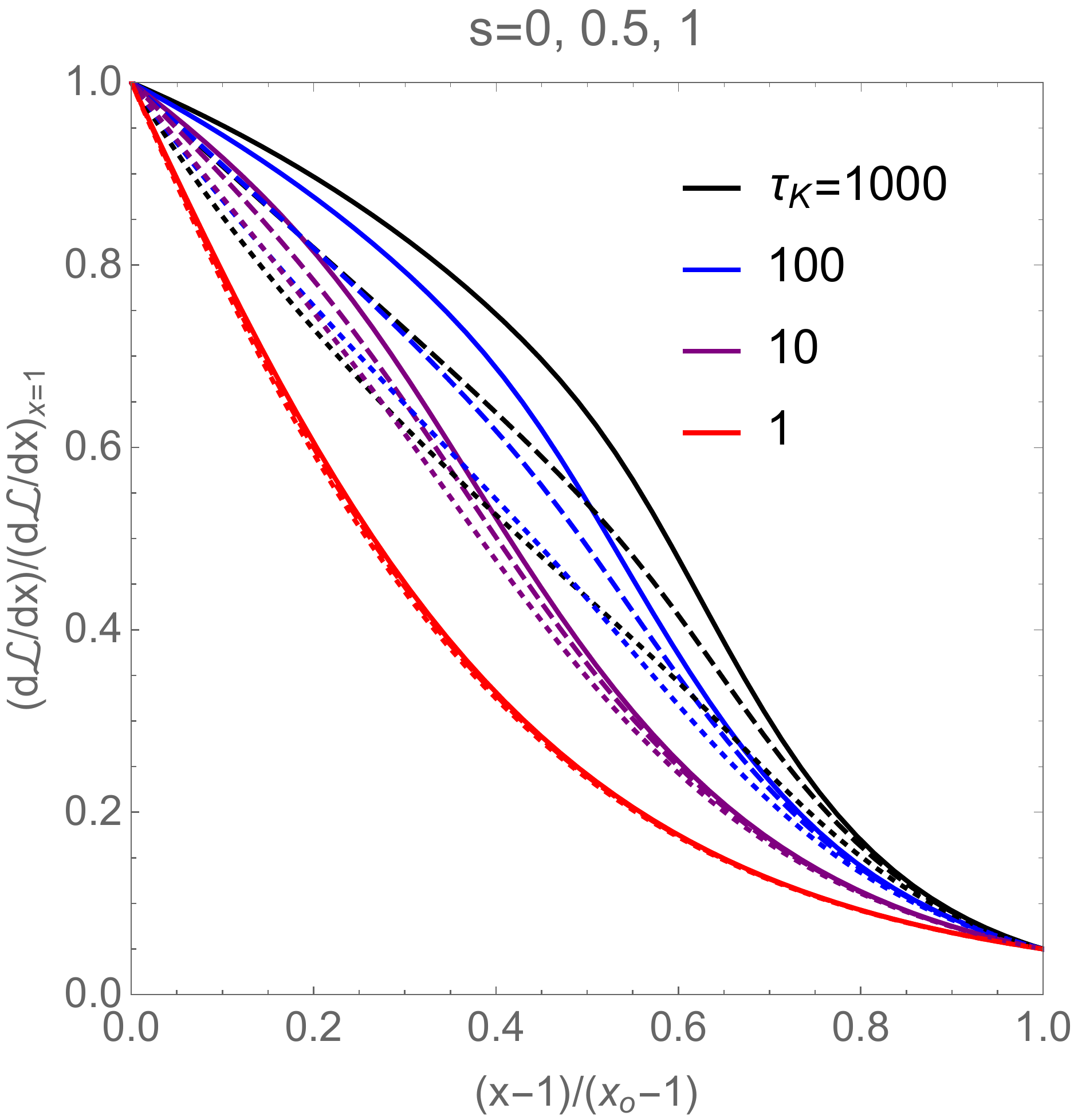}
\vskip 0.1in 
\includegraphics[scale=.25]{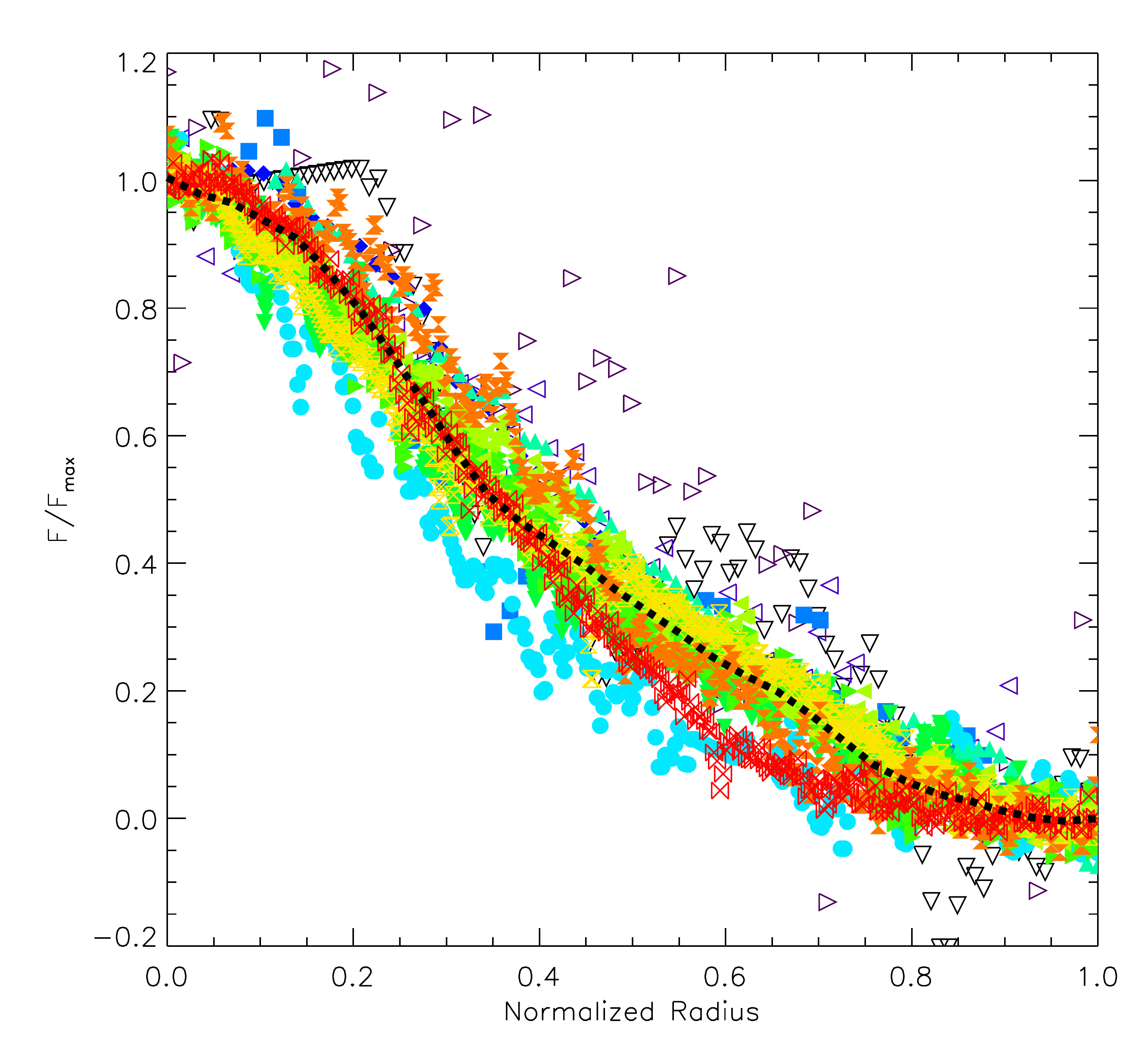}
\end{center}
\caption{
{\em Top:} For the same combination of parameters $\tau_{\rm K}$ and $s$ used in figures \ref{fig:fig5} and \ref{fig:fig6},  
the emission normalised at $x=1$ plotted vs. a normalised frequency $(x-1)/(x_{\rm o}-1)$, where 
$x_o$ is defined to have a normalized intensity of $0.05$.
{\em Bottom:} Reproduction of similarly normalised observed line profiles from figure 10 of \citet{Shultz20}.
}
\label{fig:fig7}
\end{figure}

For the same rotation and tilt assumed for figure \ref{fig:fig4}, the top row of figure \ref{fig:fig6} now compares this $I(x,y)$ for the optically thick case $\tau_{\rm K} =1000$ of an isothermal disk ($s=0$; left panel), to disks with the same $\tau_{\rm K}=1000$, but a 
temperature that declines  radially with power indices $s=1/2$ (middle panel) and $s=1$ (right panel).
 By applying the resulting $I_{\rm o} (x,y)$ into the line profile integral (\ref{eq:Lx}), the lower panel of  figure \ref{fig:fig6} shows the associated wing profile for various combinations of $\tau_{\rm K}$ and $s$.
 
 Note that increasing the power index $s$ leads to flatter profiles, with reduced or even no downward  concavity.
To see the effect of both increasing index $s$ and decreasing optical depth, the upper panel of figure \ref{fig:fig7} compares $d\mathcal{L}/dx$ normalised by its value at $x=1$, now plotted vs. a  normalised frequency variable $(x-1)/(1-x_{\rm o})$, wherein $x_{\rm o}$ is defined to be an outer frequency at which $(d\mathcal{L}/dx) (x_{\rm o}) \equiv 0.05 (d\mathcal{L}/dx) (1) $.
 
 The lower panel of figure \ref{fig:fig7} reproduces the similarly normalised observed emission profiles given in figure 10 from \citet{Shultz20}.
 The dark dots for the average to the scattered data points do show a notable downward concavity in the inner wing, indicating that many of the disks sampled must have a nearly constant surface brightness in the optically thick region near and above the Kepler radius.
 This in turn suggests at least some disks in the observed sample must be both moderately optically thick and have a temperature that does not drop too steeply in radius, i.e. $s<1$.

\begin{figure}
\begin{center}
\includegraphics[scale=0.45]{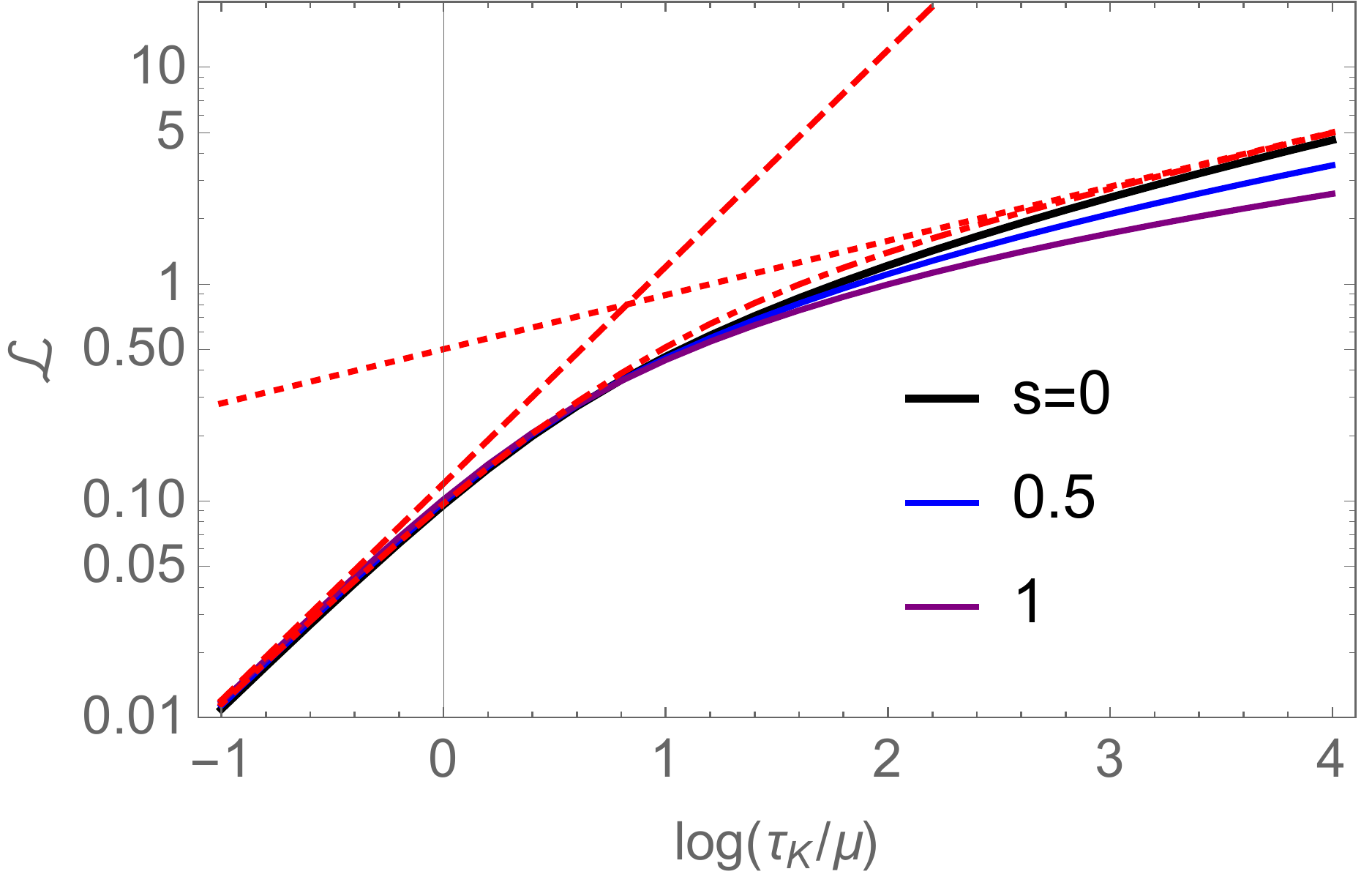}
\caption{For standard tilt ($\mu=1/\sqrt{2}$) and rotation ($W=1/2$) parameters, the growth of disk-integrated emission with Kepler optical depth $\tau_{\rm K}$, comparing cases with source function power-law $s=0$, 0.5, and 1. 
The dashed and dotted red lines show fits for the optically thin and thick limits, $\mathcal{L} \sim \tau_{\rm K}$ and 
$\mathcal{L} \sim \tau_{\rm K}^{1/4}$, while the dashed-dotted red line shows the simple analytic fit given by equation (\ref{eq:Lfit}).
Since $\mathcal{L}$ is proportional to the emission equivalent width (see equation (\ref{eq:EW})), this plot represents a ``curve of growth'' for disk emission.
}
\label{fig:fig8}
\end{center}
\end{figure}

\subsection{Curve of growth for emission equivalent width}

The overall disk emission from the line wing $x>1$ is given by the integral,
\beq
\mathcal{L} \equiv  \int_1^\infty \frac{d\mathcal{L}}{dx} \, dx
\, .
\label{eq:ell} 
\eeq
For the same fixed tilt $\mu=1/\sqrt{2}$ and rotation parameter $W=1/2$, figure \ref{fig:fig8} plots the increase in $\mathcal{L}$ with the Kepler optical depth $\tau_{\rm K}$, comparing results for source function power indices $s=0$, 0.5, and 1.
In the optically thin domain $\tau_{\rm K} < 1$, the emission increases linearly with optical depth, as shown by the dashed line fit,
$ \mathcal{L} = 0.12 \, \tau_{\rm K}/\mu$.
But in the optically thick range here, we find\footnote{In the strong optically thick limit, we expect $\mathcal{L} \sim R_{\rm o}^2$, where the outer disk radius $R_{\rm o} \approx (\tau_{\rm K}/\mu)^{1/12}$, implying $\mathcal{L} \sim ( \tau_{\rm K}/\mu)^{1/6}$. But we find this scaling is only reached asymptotically for $R_{\rm o} \sim 10$, and so extreme optical thickness $\tau_{\rm K} \sim 10^{12}$.}
$ \mathcal{L} \sim \tau_{\rm K}^{1/4}$.
The dot-dashed red curve shows a simple form that bridges these two limits,
\beq
\mathcal{L}  \approx \frac{0.12}{\mu/\tau_{\rm K} + 0.24 \, (\mu/\tau_{\rm K})^{1/4}}
\, ,
\label{eq:Lfit}
\eeq
which as shown provides a very good fit to the full $\mathcal{L}$ result  for an isothermal ($s=0$) disk (black curve).

This dimensionless emission can be converted to a physical luminosity for disk limb emission from {\em both} profile wings, $L= 2 \mu S_{\rm K} R_{\rm K}^2 \mathcal{L}$.  
By comparison,  a star with surface brightness $I_\ast$ and radius $R_\ast$ has a luminosity $L_\ast = I_\ast \pi R_\ast^2$.
Dividing by this and multiplying by the thermal Doppler width $\Delta \lambda_{\rm D} $ associated with integration over the profile function $\phi_\lambda$, we obtain the associated emission equivalent width,
\beq
\boxed{
W_\lambda =
\frac{2 \mu}{\pi} \,
\frac{ R_{\rm K}^2 }{\ R_\ast^2} \,
\frac{S_{\rm K}}{I_\ast} \,
\Delta \lambda_{\rm DK}  \,
\mathcal{L}
\, .
}
\label{eq:EW}
\eeq
Since $\mathcal{L} \sim W_\lambda$, figure \ref{fig:fig8} thus represents a ``curve of growth'' for disk emission with $\tau_{\rm K}$.

Moreover, in terms of the Kepler-radius field strength $B_{\rm K}$ and its critical value $B_{\rm K1}$ for unit optical depth from equation (\ref{eq:BK1b}), 
we see that  $\tau_{\rm K} = (B_{\rm K}/B_{\rm K1})^4 \equiv b_{\rm K}^4$, and thus that this curve of growth can be equivalently cast in terms of the increase of $\mathcal{L}$ with the Kepler field strength $B_{\rm K}$.
In particular, this can be implemented through the fit form (\ref{eq:Lfit}) with the simple substitution $\tau_{\rm K} \rightarrow b_{\rm K}^4$.

\section{Comparison with Observations}

Let us now make direct comparisons of these theoretical scalings with the observed emission properties of the sample of magnetic B-stars analysed by \citet{Shultz20}.

For this, we first note
that our theoretical analysis has only explicitly considered the idealised special case of rotation-aligned dipoles, whereas the actual stars are inferred to have rotation-field tilt angles over the full range $ 0 < \beta < 90^o$, as well as non-dipole components.
Such tilted or non-dipolar fields break the axi-symmetry of the simple rotation-aligned case, leading to a rotational modulation of the observed H$\alpha$ emission, and even extra absorption when circumstellar clouds occult the star.
Moreover, their accumulation surfaces  are no longer a planar disk at the common rotational and magnetic equator, but 
instead develop a warped form with increasing field-rotation tilt angle $\beta$.

Appendix A of \citet{Townsend08} showed, however, that is this accumulation surface has a mean normal that has only a moderate tilt angle relative to the magnetic axis,
\beq
{\bar \nu} \approx - \arctan \left ( \frac{\sin 2 \beta}{5 + \cos 2\beta} \right )
\, .
\label{eq:nubar}
\eeq
This  is zero at both $\beta = 0$ and 90$^o$, and has an extremum of just ${\bar \nu} \approx -11^o$ at $\beta \approx 51^o$.

To minimise the complexity from these inherently 3D effects,  \citet{Shultz20}  focused on the rotational phase of {\em maximum}  emission in the line wings formed at projected distances beyond the Kepler radius.
This maximum occurs when the line-of-sight projection of the warped disk is greatest, as computed using (\ref{eq:nubar}). 
By dividing the observed maximum emission by this maximum projection,
 the observations can be more appropriately compared with predictions of the simplified aligned-dipole model for  CBO. 

\subsection{Critical field for onset of detectable emission}  

As noted in the introduction, a key motivation for the theoretical analysis in this paper was the result   (shown in the right panel of figure 3 from \citet{Shultz20}) that the {\em onset} of detectable H$\alpha$ occurs at critical value of the Kepler radius field strength, $B_{\rm K} \approx 100$\,G.

Indeed, \citet{Shultz20} find even cleaner transition for the wind-mass-loss-corrected ratio of the Alfven radius to Kepler radius, which has the scaling $R_{\rm A}/(R_{\rm K} {\dot M}^{1/4})  \sim B_{\rm K} R_{\rm K}$ (see the middle panel of their figure 3).
Comparing with equation (\ref{eq:BK1a}) here, and noting that $\sqrt{g_{\rm K}} \sim 1/R_{\rm K}$, we see that this transition scaling
 is even closer
 in matching the predicted the analytic scaling for $B_{\rm K1}$.

\begin{figure}
\begin{center}
\includegraphics[scale=0.5]{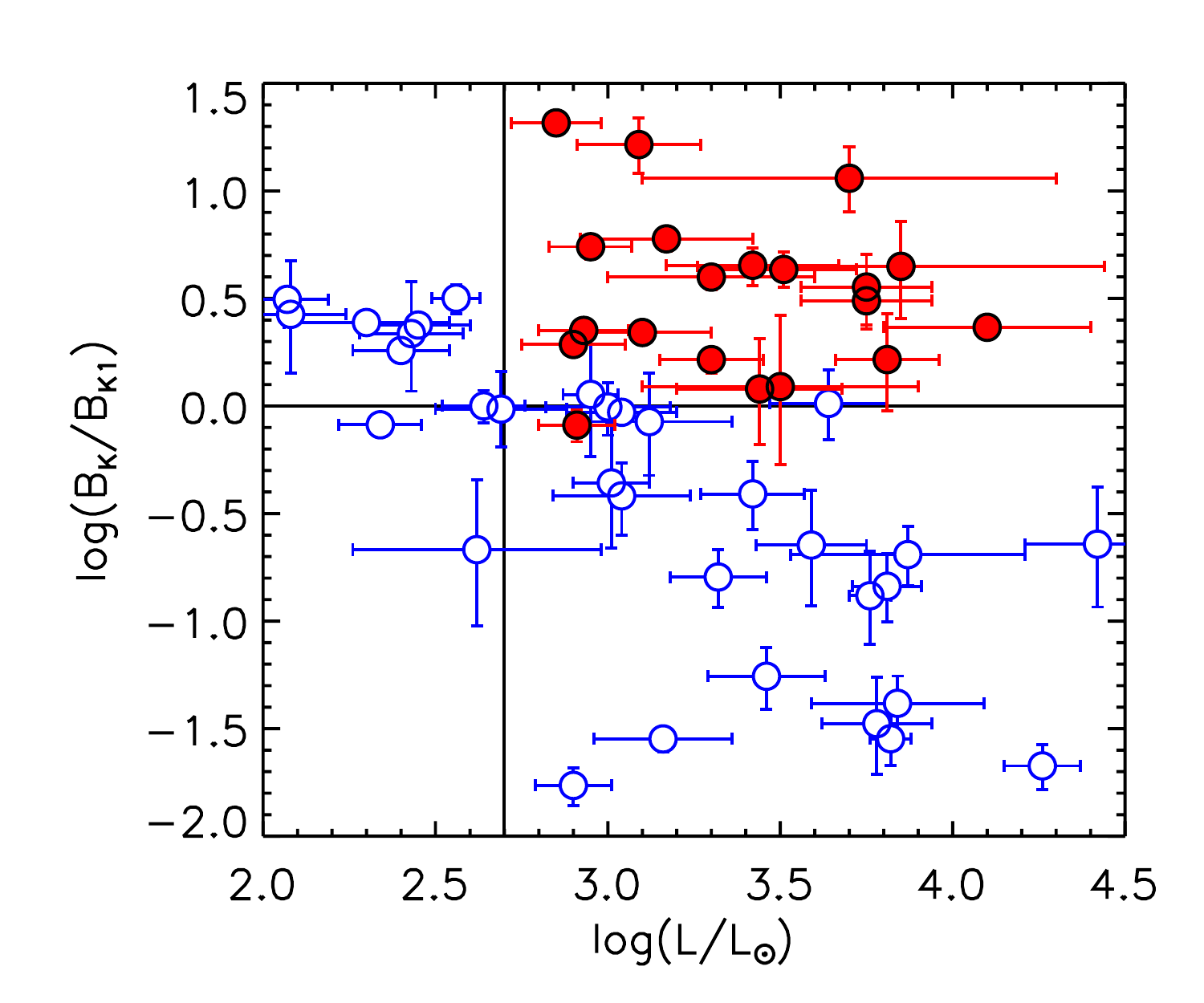}
\includegraphics[scale=0.5]{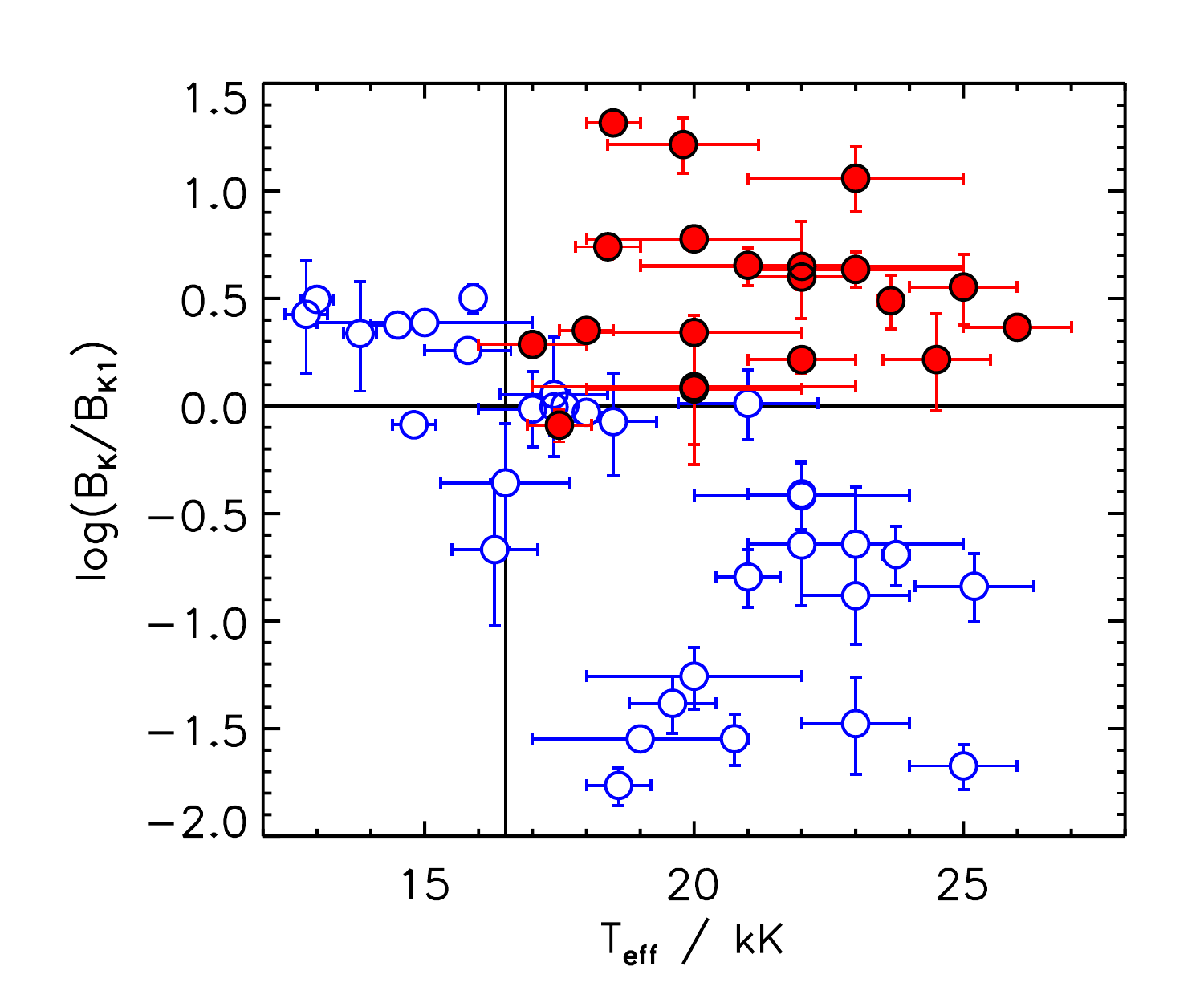}
\caption{
{\em Top}:
Observed magnetic B-stars plotted in the $\log B_{\rm K}/B_{\rm K1} $ vs.\ $\log L/L_\odot$ plane, with stars showing H$\alpha$ emission in filled red circles, and those without emission in open blue circles. 
{\em Bottom}: Same as top, but plotted vs.\  log of stellar effective temperature (in K).
In both panels, the vertical dashed line marks the empirical lower limits in luminosity or effective temperature for even strongly magnetic stars to show Balmer emission.
}
\label{fig:fig9}
\end{center}
\end{figure}

Choosing the {\em average} between the nebular and LTE opacities, and assuming a constant disk temperature $T_{\rm K} \approx T_{\rm eff}$, we thus apply the inferred values for rotation period $P$ and Kepler gravity $g_{\rm K}$ to derive for each star in our sample values for $B_{\rm K1}$, and associated values for the ratio $b_{\rm K} \equiv B_{\rm K}/B_{\rm K1}$.
Figure \ref{fig:fig9} here then plots the stars in the $\log b_{\rm K} $ vs.\ $\log L$ (top panel) and $\log b_{\rm K} $ vs.\ $\log T_{\rm eff}$ (bottom panel) planes, again marking those with detectable emission with filled red circles, and those without with open blue circles.
Note that the horizontal line at $b_{\rm K}=1$ does remarkably well at separating the stars with and without detectable emission,
as well as or better than the empirical separations in figure 3 of  \citet{Shultz20}.

This represents strong evidence that centrifugal breakout is the mechanism controlling the mass loss from the CMs in these early to mid-B type magnetic stars.

\subsection{Lack of emission in late-B and A stars}

Figure \ref{fig:fig9} also shows, however, that there is a limit in luminosity or effective temperature below which even strongly magnetic 
stars\footnote{As in the right panel of figure 3 from \citet{Shultz20}, the open squares represent cooler, lower luminosity, late-B and A-type stars.
They were added to our original sample of early- to mid-B stars to illustrate further this cool-star cutoff in detectable emission.} 
{\em above} the horizontal line for $B_{\rm K}=B_{\rm K1}$ show {\em no} detectable emission. These appear
to the left of the {\em vertical} dashed lines at $\log (L/L_\odot) = 2.8$ and $\log (T_{\rm eff}/K) = 4.22$ ($T_{\rm eff} = 16.6$\.kK), corresponding to spectral type B6. 

This is likely associated with the sharp drop in radiatively driven stellar wind mass loss rates for such lower luminosity stars
\citep{Vink01,Krticka14}.
It suggests that there may be another competing mechanism for mass leakage from these CMs, perhaps associated with the drift and diffusion processes discussed by \citet{Owocki18}.
For low enough feeding rate from the stellar wind, such residual leakage would prevent the mass in the CM from building up to the level needed for the disk to become optically thick, and thus have an effective emission area that competes with that from the star.
From equation (\ref{eq:taur}), the surface density needed to make the Kepler radius optically thick is
\beq
\sigma_1 = \sqrt{\frac{2 \pi c_{\rm s}}{\Omega C_{\rm o}}}
\, ,
\label{eq:sig1}
\eeq
where we have used equation (\ref{eq:hK}) for the Kepler radius scale height $h_{\rm K}$.

To place a constraint on leakage, let us compare this with equation (14) from \citet{Owocki18} for the characteristic surface density at the Kepler radius from the net {\em drift} leakage against a competing wind feeding rate ${\dot \sigma}_{\rm K} \equiv {\dot M} R_\ast/(4 \pi R_{\rm K}^3)$,
\beq
\sigma_{\rm dn} = \frac{{\dot \sigma}_{\rm K}}{\Omega^2 \, t_{\rm K}}
\, ,
\label{eq:sigdm}
\eeq
where $\Omega=2 \pi/P$ is the stellar rotation frequency, and we have used the slightly modified notation $t_{\rm K}$ to represent the characteristic drift time ($\tau_{\rm K}$ in their notation).
Setting $\sigma_{\rm dn}=\sigma_1$, we can solve for a critical drift time for unit optical depth,
\beqa
t_{\rm K1} &=& \frac{{\dot M}}{4 \pi R_\ast^2} \, \frac{R_\ast^3}{R_{\rm K}^3} \, \sqrt{\frac{C_{\rm o}}{2 \pi c_{\rm s} \Omega^3}}
\\
&=& \frac{{\dot M}}{4 \pi v_{\rm orb}^2} \,  \sqrt{\frac{ C_{\rm o}}{ c_{\rm s} P}}
\\
&=&
C_{\rm K1}
~ {\dot M}_{-11} \,  \frac{R_\ast}{R_\odot} \, \frac{M_\odot}{M_\ast} T_{\rm 20kK}^{-q} \, P_{\rm day}^{-1/2}
\, ,
\label{eq:tK1}
\eeqa
where $ {\dot M}_{-11} \equiv ({\dot M}/10^{-11} \, M_\odot/$yr). 
For the LTE model, the coefficient 
$C_{\rm K1} = 2.1$\,s 
and the power index $q \approx 2.2$, while 
for the nebular model, 
$C_{\rm K1} = 0. 38$\,s 
and $q \approx 1$.
For drift times shorter than this critical value, $t_{\rm K} <  t_{\rm K1}$, the disk should remain optically thin, even for $B_{\rm K} > B_{\rm K1}$.
Taking the stars at the vertical dashed transition to have a mass $M \approx 6 M_\odot$, then using the associated values for luminosity and effective temperature in the \citet{Vink01} mass loss scaling formula (25), we obtain ${\dot M}_{-11} \approx 2$. If accurate, this would imply a leakage time  in the range 
$t_{\rm K}= 0.8$ to $4$\,s.
This is about 1.5-2 dex shorter than the very rough estimate for drift timescale $t_{\rm K} \approx 120$\,s given in section 4.2 of \citet{Owocki18}.

Alternatively, we note that such a mass loss rate is near the value expected for the onset of {\em ion runaway} \citep[see eq. 23 of][]{Owocki02}. This occurs when the wind density becomes so low that heavy minor ions that line-scatter stellar radiation are no longer well coupled by Coulomb collisions to the protons
\citep{Krticka00,Krticka01}.
Since the resulting {\em metal ion wind} thus lacks the hydrogen central to H$\alpha$ emission,  a transition to ion runaway could be the key to the observed lack of H$\alpha$ emission for stars with lower luminosity and effective temperature than the critical values marked by the vertical dashed lines in figure~\ref{fig:fig9}.

To summarise, this transition to no emission in magnetic stars with $\log L/L_\odot < 2.8$ (or $\log T_{\rm eff}/K < 4.22 $) could either provide a diagnostic for residual leakage by drift or diffusion,
or alternatively for an ion runaway transition to a metal ion wind without the Hydrogen needed for H$\alpha$ emission.

\subsection{Comparison with observed emission equivalent width}

Let us finally compare the predicted scalings  for emission {\em  equivalent width} with corresponding observational results for this sample.

The dimensionless equivalent width $\mathcal{L}$ derived in section 3.2 depends on the optical thickness at the Kepler radius $\tau_{\rm K}$, which in turns depends on the ratio of the magnetic field strength there, $B_{\rm K}$, to the critical value $B_{\rm K1}$ for unit optical depth.
As given by equation (\ref{eq:BK1a}), the latter depends on the rotation period $P$, and the temperature $T_{\rm K}$ and gravity $g_{\rm K}$ at the Kepler radius.
From equation (\ref{eq:EW}) conversion to dimensional form depends on the inclination $\mu$, the Kepler values of the radius $R_{\rm K}$, and the temperature-dependent source function $S_{\rm K}$ and thermal Doppler width $\lambda_{\rm DK}$.

\begin{figure}
\begin{center}
\includegraphics[scale=0.4]{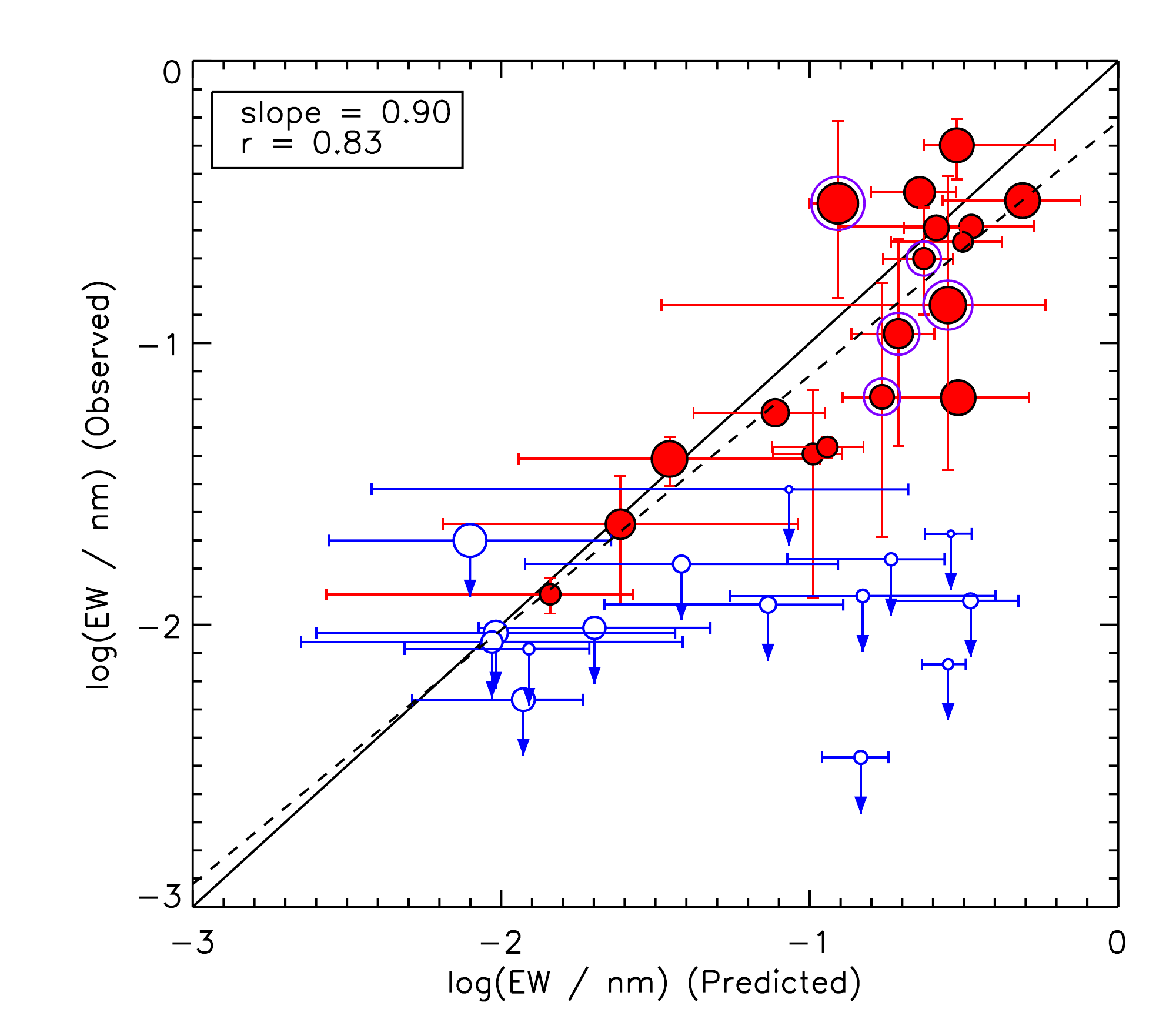}
\includegraphics[scale=0.4]{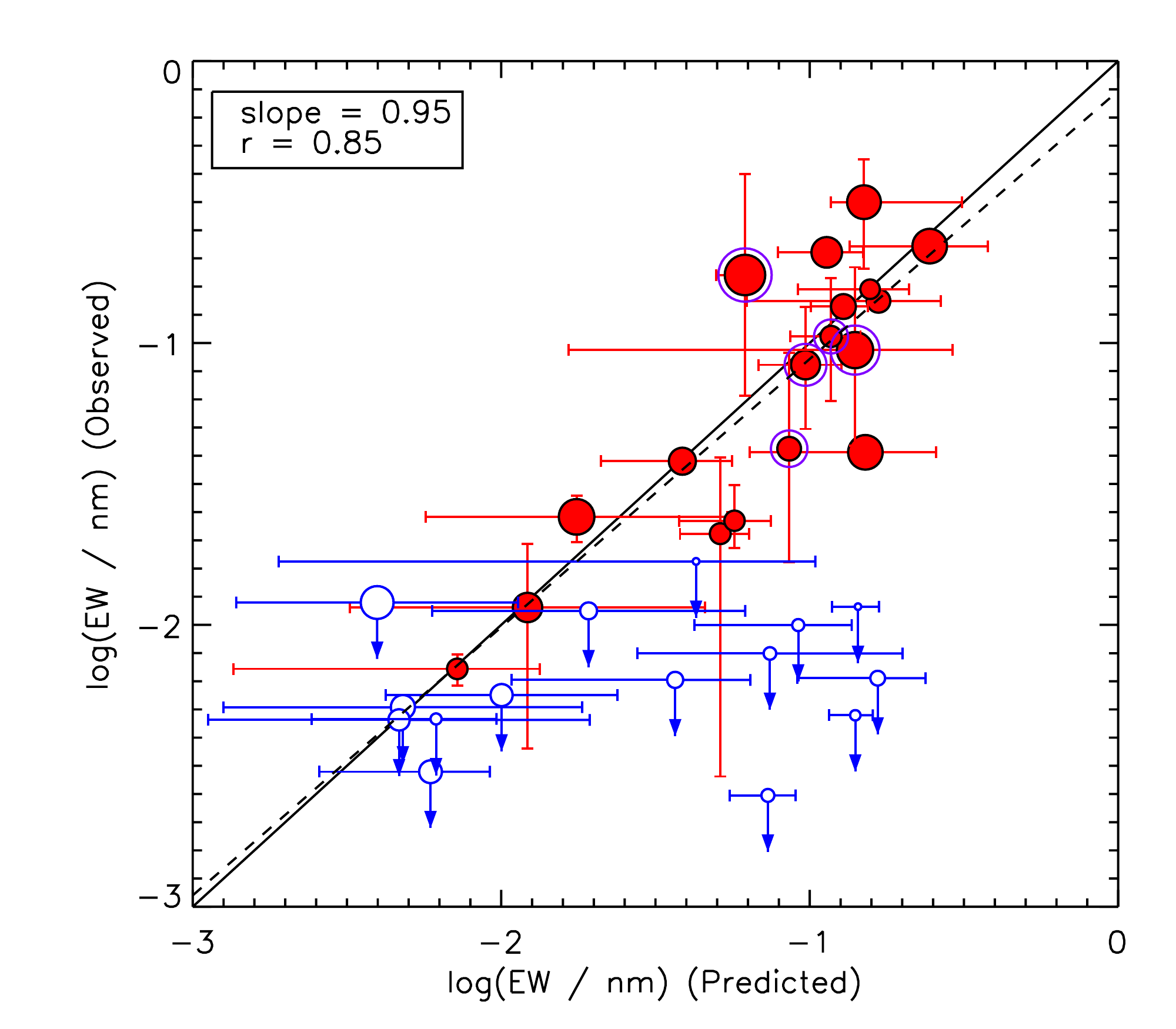}
\caption{For the magnetic B-star sample from \citet{Shultz20}, measured equivalent width (EW) vs.\ predicted values from the CBO model, plotted on a log-log scale.
Symbol size scales with log luminosity, with open blue circles showing upper limit of EW of stars without detectable emission.
Red filled circles with a blue ring mark spectroscopic binaries, highlighted because there is an additional degree of uncertainty in their EW (though it has been adjusted  for dilution).
The top panel gives results including both profile wings, but because binarity can affect the symmetry of the profile, the bottom panel instead uses the EW from the wing less affected by the companion.
For the latter, the derived slope is indeed marginally closer to the expected unit value, with also a somewhat better linear regression coefficient $r$.
The dashed lines show this best-fit linear regression, while the solid lines show the direct relation, i.e., measured = predicted.
}
\label{fig:fig10}
\end{center}
\end{figure}

For an isothermal disk with again $T_{\rm K} \approx 0.75 T_{\rm eff}$, 
and thus $S_{\rm K}= 0.75 I_\ast$,
figure \ref{fig:fig10} plots on a log-log scale the observed vs.\ predicted equivalent width $W_\lambda$.
The top panel includes both wings, but to account for the effects of a non-magnetic binary companion (whose systems are flagged with large red circles), 
the bottom panel shows results from  the EW of just the less-affected wing.
In both panels, the solid lines signifying ``measured = predicted'' gives a good apparent fit to the data,
but the best-fit linear regressions (dashed lines) show slightly higher correlation for the one- vs. two-wing calculation, with also a slope that is closer to the expected unit value.

Overall, this quite good agreement between the measured vs. predicted EW values for this B-star sample provides further strong evidence
in favor of the CBO model
for the maximum confined mass in their CMs and its associated H$\alpha$ emission.

\section{Discussion}


\subsection{CBO scalings for Balmer emission}

The CBO analysis here
{\em predicts} a critical field strength at the Kepler radius, $B_{\rm K1}$ (given by equation (\ref{eq:BK1a})), at which the disk becomes optically thick in H$\alpha$, and thus capable of emitting over a surface area that competes with the star.
The quite remarkable, quantitative agreement of this prediction with the onset of detectable H$\alpha$ emission, as illustrated in figure \ref{fig:fig9}, provides strong evidence in support of the basic CBO paradigm.

In addition also to the strong correlation between measured and predicted equivalent width in figure \ref{fig:fig10}, the CBO model can also reproduce well the observed line profile shapes. In particular, figure \ref{fig:fig7} shows that the commonly observed downward concavity of the inner wing requires the uniform surface brightness associated with a nearly {\em isothermal} disk.
This is a natural outcome of photoionization heating, which tends to fix the temperature of circumstellar material near the stellar effective temperature (as occurs, e.g., in HII regions over distances well away from the star).

\subsection{Continuum optical thickness}

A key point in this regard is that, even though the CM disk can become very optically thick near line center of H$\alpha$, its {\em continuum} optical thickness is generally much less.
Because of uncertainty in the mechanism(s) for disk mass loss, the RRM model of CMs developed by \citet{Townsend05} is not able to predict {\em a priori} their continuum optical thickness, and so instead has relied on empirical constraints, e.g. associated with polarization \citep{Carciofi13}, or broadband occultation of the star by CM clouds \citep{Townsend13}.

The CBO model for the maximum confined mass and disk  surface density now also provides a direct means to predict {\em also} the associated {\em continuum} optical thickness. For example, for simple electron scattering opacity $\kappa_{\rm e} = 0.34$\,cm$^{2}$\,g$^{-1}$, the predicted optical thickness of the CM for radii $r>R_{\rm K}$ scales as
\beq
\tau_{\rm e} (r) \approx  0.3 \, \kappa_{\rm e} \sigma_\ast \left ( \frac{r}{R_{\rm K}} \right )^{-6}
\, .
\label{eq:taue}
\eeq
In addition to giving a fixed value at the Kepler radius, $\tau_{\rm e} (R_{\rm K}) = 0.3 \kappa_{\rm e} \sigma_\ast$, this invokes the radial decline inferred from the MHD calibration, $r^{-6}$.
As noted, this is much steeper than the $r^{-3}$ decline assumed in the RRM model, based on the notion of a fixed filling time from the wind surface feeding rate 
${\dot \sigma}  \sim B \sim r^{-3}$ \citep[see][eq.\ 34]{Townsend05}.

Applying the scaling (\ref{eq:sigk}) for $\sigma_\ast$, we find the critical field for the Kepler radius to have unit optical depth in electron scattering is
\beq
\boxed{
B_{\rm K1e} = \sqrt{ \frac{ 4 \pi g_{\rm K}}{0.3 \kappa_{\rm e}} } \approx 1110\, {\rm G} \, \sqrt{g_{\rm K4}}
}
\, ,
\label{eq:BK1e}
\eeq
where again $g_{\rm K4} \equiv g_{\rm K}/(10^4$\,cm\,s$^{-2}$).
Comparison with equation (\ref{eq:BK1b}) shows that this is just about an order magnitude higher than the critical field $B_{\rm K1}$ to make the disk have $\tau_{\rm K}=1$ at H$\alpha$ line center.
But whereas the H$\alpha$ optical depth increases with $\sigma^2 \sim B_{\rm K}^4$, the density-independent nature of electron opacity $\kappa_{\rm e}$ means its associated optical depth scales as $\tau_{\rm e} \sim \sigma \sim (B_{\rm K}/B_{K1e})^2$.
The upshot is the only CMs that are very optically thick in H$\alpha$, with $\tau_{\rm K} > 10^4$, are likely to become optically thick in electron scattering.

Equation (\ref{eq:BK1e}) can alternatively be cast in terms of an equatorial {\em surface} field needed to make the disk optically thick,
\beq 
B_{\ast } [\tau_{e} (R_{\rm K}) =1] \approx 1110 \, {\rm G} \, \sqrt{g_{4}} \left ( \frac{R_{\rm K}}{R_\ast} \right )^2
\, ,
\label{eq:Bse}
\eeq
where now $g_{4} \equiv  g_\ast/(10^4$\,cm\,s$^{-2}$). 
The required polar field is just twice this equatorial surface value.

\subsection{Total mass of CM}

The total mass in the CM can be computed from integrating outward from the Kepler radius,
\beqa
M_{\rm CM} &\equiv& \int_{R_{\rm K}}^\infty  \sigma(r) \,  2 \pi r \, dr 
= 
0.6 \pi \sigma_\ast  \int_{R_{\rm K}}^\infty \left ( \frac{r}{R_{\rm K}} \right )^{-6} r \, dr
\nonumber
\\
&=& 0.6 \frac{B_{\rm K}^2 R_{\rm K}^2}{16  g_{\rm K}}
=
0.0375 \frac{B_\ast^2 R_\ast^4}{  g_\ast R_{\rm K}^2}
\, .
\label{eq:MCM}
\eeqa
By comparison, equation (A11) of \citet{Townsend05} derived a similar expression for the asymptotic disk mass $m_\infty$, which has the same parameter scaling, but which is about a factor 8 ($\approx \sqrt{\pi}/6/0.0375$) higher than the expression (\ref{eq:MCM}) derived here.
Moreover, in their example application of this scaling for the parameters inferred for the prototypical CM star $\sigma$\,Ori\,E, \citet{Townsend05} quote a total mass $m_\infty = 9.4 \times 10^{-8} M_\odot $.
But in deriving this value, they erroneously applied the inferred {\em polar} value for the surface field, $B_{\rm p} \approx 10^4$\,G, whereas the magnetic confinement in their breakout analysis is set by the  field strength at the magnetic {\em equator}, $B_\ast = B_{\rm p}/2$. Since $m_\infty \sim B_\ast^2$, this makes their quoted value a further factor 4 too high.

Overall, the CM estimates from TO05 are thus roughly a factor 32, or about 1.5 dex, higher than our new scaling (\ref{eq:MCM}) based on a MHD-calibrated model for CBO.

\subsection{Counters to arguments against breakout}

Such prior overestimates of CM mass from breakout are significant because a key critique against the CBO scenario later raised by \citet{Townsend13} was that the CM mass they inferred from photometric absorption by CM clouds was much less than the putative values inferred from the breakout analysis in Appendix A of \citet{Townsend05}.
Specifically, for their assumed parameters for $\sigma$\,Ori\,E, -- viz. $M_\star  = 8.3 M_\odot$, $R_\ast = 3.77 R_\odot$,  $R_{\rm K} = 2.54 R_\ast$, $B_{\rm p} = 11,000$\,G -- they inferred an asymptotic breakout mass $m_\infty  \approx 1.2 \times 10^{-8} M_\odot$, nearly two orders of magnitude higher than their empirically estimated upper limit, $2 \times 10^{-10} M_\odot$.

By comparison, if we apply these same parameters in our equation (\ref{eq:MCM}), we obtain for the total CM mass $M_{\rm CM} \approx 3.8 \times 10^{-10} M_\odot$, which is now within a factor two of their empirically inferred value.

The associated normal optical thickness for electron scattering at the Kepler radius is $\tau_{Ke} \approx 0.37$.
Along a line of sight with projection cosine $\mu = \cos {\bar \nu} $ to the disk normal the optical thickness increases by a factor $1/\mu$.
This can allow the occultation of the disk against the star to give notable photometric dips in the continuum, as observed for $\sigma$\,Ori\,E, wherein such dips are found to have a minimum about 10\% of the continuum.

The upshot here is that the basic scaling for surface density in our MHD-calibrated breakout analysis,
grounded by equation (\ref{eq:sigk}) (which is based on equation (A4) from \citet{Townsend05}), is in fact quite compatible with empirical inferences based on photometric absorption, as well as the observed level of polarization \citep{Carciofi13}.

\citet{Townsend13}, have also cited the steady repeatability of the photometric variation of $\sigma$\,Ori\,E as an argument against breakout, since this shows no evidence for the strong disruption seen in 2D simulations of CBO events, e.g. by \citet{udDoula09}.
But in 3D simulations \citep{udDoula13}, such breakouts exhibit a strong azimuthal incoherence, with also a hierarchy of breakout amplitudes;  when volume-averaged over azimuth, the level of variability is greatly reduced, allowing then for the nearly steady repeatability in the photometric light curve.
Moreover, while there were initial suggestions that breakout events might explain X-ray emission and flares \citep{udDoula06}, similar arguments about 3D global averaging can also explain the lack of clear X-ray flaring in magnetic B stars.

\subsection{Limitations}

Let us finally put these successes of our CBO analysis into context.
A key limitation stems from the idealised restriction to rotation-aligned dipole, whereas most all the sampled magnetic B-stars are generally inferred to have a non-zero tilt angle between the magnetic and rotation axes, $0 < \beta < 90^o$, as well as in some cases significant non-dipole field components.
This breaks the axisymmetry of the idealised aligned-dipole model, leading to a warped accumulation surface, with the highest density occurring at azimuths with Kepler offset at the smallest radius.
Although the resulting mass distribution is inherently 3D, our 2D field-saligned analysis of CBO seems still to model well the {\em peak} of rotationally modulated emission in the line wings; this arises from the accumulation disk above the Kepler radius when it appears outside the stellar limb with maximum projected area.

But it still unclear how to quantify the level and radial decline of the mass surface density at azimuths with lower density and weaker confinement.
For this there is a need to account for the inherent tilt between the confining magnetic field normal and the outward net centrifugal force that is directed away from the rotation axis.
The greater geometric complexity, and the dependence on latitude and azimuth will make an extension of the present analytic approach difficult, and so in need even more urgently of calibration with full 3D MHD simulations of such tilted dipole cases.  This is an area of current focus in our group.

In this context, we note that the clear demonstration here that CBO is the primary mechanism for controlling the level and distribution of the CM mass implies also inherent limitations for both the RRM model of CMs, as well as the associated `Rigid Field Hydro-Dynamics' \citep[RFHD;][]{Townsend07} approach for simulating the mass accumulation in such CMs.
Central to both approaches is the notion that the magnetic field is so strong that it effectively acts as a completely {\em rigid} conduit that both channels the outflowing stellar wind, and confines the resulting CM material against the net outward centrifugal force.

But the result here that CBO controls the 
 maximum confined mass of a CM
implies a complex, dynamical distortion of the field that is intrinsic to observed CMs and their emission.
This, along with the steeper than assumed radial decline in surface density ($r^{-6}$ vs.\ $r^{-3}$), could be a key factor in the lingering discrepancies  \citep{Oksala12,Oksala15b} that the RRM and RFHD models show in reproducing many details of the observed dynamic spectrum of H$\alpha$ for $\sigma$\,Ori\,E (even when inferred non-dipole components of the stellar field are accounted for through the `arbitrary' field forms of the RRM and RFHD models).
Moreover, the neglect of such non-dipole components represents also a further limitation for the efficacy of our CBO analysis.

Finally, both this CBO analysis and the RRM/RFHD models envision the peak CM density occurring very near the Kepler radius.
But the sample of observed B-star CMs with Balmer emission show distinct evidence for the peak emission occurring at wing frequencies well above that associated with $R_{\rm K}$, typically by a factor $1.4$ but ranging even up to nearly a factor $2$ \citep[see figure 7 of][]{Shultz20}.
This suggests there could be additional processes controlling the filling and confinement of material in the innermost part of the CM (e.g., perhaps leakage back down to the star).
In any case, it represents a key unsolved puzzle and thus a clear limitation for both the CBO and RRM paradigms for modeling CMs.

\section{Summary and Future Work}

Let us conclude with an itemised summary of the analysis and results of this paper.

\begin{itemize}
\item Motivated by the empirical discovery by \citet{Shultz20} that the onset of detectable H$\alpha$ emission from centrifugal magnetospheres (CMs) in magnetic B-stars is {\em independent} of stellar luminosity, and thus of the stellar wind feeding rate of the CM, we have reexamined \citep{Townsend05} here the notion that the eventual loss of CM mass occurs through {\em centrifugal breakout} (CBO).
\item  Our CBO analysis predicts a quantitative scaling for the level and radial distribution of CM surface density $\sigma(r)$ (equations (\ref{eq:sigk}) and (\ref{eq:sigr})).
\item The associated optical depth in H$\alpha$ scales with Kepler radius values $\tau_{\rm K} \sim \sigma(R_{\rm K})^2 \sim B_{\rm K}^4$, with thus a sudden onset of detectable emission at a critical Kepler field $B_{\rm K1}$ (equation (\ref{eq:BK1a})) that gives $\tau_{\rm K} \approx 1$, as demonstrated by figure \ref{fig:fig9}.
\item The associated curve of growth of emission equivalent width with increasing optical depth $\tau_{\rm K} = (B_{\rm K}/B_{\rm K1})^4$ (figure \ref{fig:fig8}) leads to a predicted equivalent width scaling (equations (\ref{eq:Lfit}) and (\ref{eq:EW})) that matches well the observed values (figure \ref{fig:fig10}).
\item This CBO model can also reproduce the commonly observed downward concavity of the line-profile inner-wing, but only if the CM is nearly isothermal (figure  \ref{fig:fig7}), and
at least moderately optically thick.
\item While the CBO model explains well the emission in early- to mid-B stars, spectral types later than about B6 (with $T_{\rm eff} \lesssim 16$\,kK and luminosity $L \lesssim 800 L_\odot$) show {\em no} emission, even for stars with $B_{\rm K} > B_{\rm K1}$. This might signify a residual diffusive/drift leakage that prevents the lower stellar wind mass loss from filling the CM to the level needed for H$\alpha$ emission, or alternatively might result from a transition to a metal-ion wind that lacks the requisite Hydrogen.
\item The total CM mass from the MHD-calibrated CBO analysis here is an order of magnitude lower than previous estimates, and is no longer incompatible with values inferred empirically by \citet{Townsend13}. This and the quasi-steady nature of breakout in 3D models  thus effectively mitigate these authors' arguments against the CBO paradigm.
\item A remaining puzzle is that this CBO model, like its RRM and RFHD predecessors, cannot explain the fact that observed line profiles show peak emission from well above the Kepler radius. 
\item To address this and other limitations, further work is needed to generalise the aligned-dipole assumption to model the CBO and resulting azimuthal mass distribution in actual stars with non-zero tilt angle $\beta$, and with non-dipole field components.
\end{itemize}

Overall though, the remarkable agreement found here between theoretical and empirical scalings seems to establish quite clearly that CBO is the key mechanism controlling the mass and emission properties of the CMs from magnetic early- to mid-B stars.
These predicted scalings could thus even provide leverage to infer the magnetic properties of stars that are too faint for direct spectropolarmetric detection of a field, but which are bright enough to detect H$\alpha$  emission and measure its associated equivalent width.
The breakout of a substantial mass, and the energy release from the associated magnetic reconnection, might also lead to additional observational signatures, e.g. in X-rays
 \citep{udDoula06}.

In conclusion, the quantitive CBO predictions developed here for CM mass and emission provide a new linchpin toward a more complete and quantitative understanding of these magnetic B-stars, and of the complex interplay between rotation and magnetic field channeling and confinement of the star's wind outflow.

\section*{Acknowledgments}
MES acknowledges the financial support provided by the European Southern Observatory studentship program in Santiago, Chile; the Natural Sciences and Engineering Research Council (NSERC) Postdoctoral Fellowship program; and the Annie Jump Cannon Fellowship, supported by the University of Delaware and endowed by the Mount Cuba Astronomical Observatory. 
This AJC fellowship played a particularly central role in 
facilitating the close interaction 
 between theory and observations that forms the basis of the analysis presented here.
 A.uD  acknowledges  support  by  NASA  through  Chandra  Award numbers GO5-16005X and TM7-18001X issued by the Chandra X-ray Observatory Center which is operated by the SmithsonianAstrophysical Observatory for and on behalf of NASA under con-tract NAS8- 03060.
 RHDT acknowledges support from the NSF through awards ACI-1663696 and AST-1716436.
 SPO acknowledges a visiting  professor  scholarship ZKD1332-00-D01 at KU Leuven and its Institute voor Sterrenkunde (IvS), which helped initiate work with JOS on Balmer emission from magnetic stars.
 Finally, we thank the referee, John Landstreet, for constructive comments and suggestions that improved the paper.

\section*{Data Availability Statement}

 No new data were generated or analysed in support of this research. All data referenced were part of the associated publication, \citet{Shultz20}.

\bibliographystyle{mn2e}
\bibliography{OwockiS}

\phantom{.}

\appendix

\section{Models for Balmer line opacity}

Let us first write the H$\alpha$ opacity in the general form,
\beq
 \rho \kappa_\nu = n_2 f_{23} \sigma_{\rm cl} \phi_\nu
\, ,
\label{eq:kaprho}
\eeq
where 
$n_2$ is the number density of H-atoms in the $n=2$ level,
$f_{23}$ is the oscillator strength for upward H$\alpha$ transition from $n=2$ to $n=3$,
and 
$\sigma_{\rm cl}  $  is the classical oscillator,
given in CGS units by  $\pi e^2 / m_{\rm e} c$, with $e$ and $m_{\rm e}$ the electron charge and mass.
Also, $\phi_\nu$ is the line-profile function at frequency $\nu$ (with dimension of inverse frequency).
An associated profile in wavelength is  $\phi_\lambda = \phi_\nu c/\lambda^2$, with thus the wavelength-dependent opacity $\kappa_\lambda = \kappa_{\rm o} \phi_\lambda$,
where the opacity at line-center wavelength $\lambda_{\rm o}$ is
\beq 
\rho \kappa_{\rm o} =  n_2 f_{23} \sigma_{\rm cl} \phi_{\lambda_{\rm o}} \lambda_{\rm o}^2/c
\eeq

To proceed, we need to estimate the number density $n_2$.
In B-star disks we expect most of the Hydrogen to be ionised into protons, with thus the total Hydrogen number density $n_H \approx n_{\rm p}$. 
The small number of neutrals at level 2 arise from recombination of the protons with electrons, with thus a dependence on the product of the number densities for electrons and protons,
\beq
n_2 = n_{\rm p} n_{\rm e} \Phi(T) 
\, ,
\eeq
where $\Phi(T)$ is temperature-dependent factor that depends on the details of the ionization/recombination equilibrium.
Below we invoke two distinct models for computing $\Phi (T)$, one based on an {\em LTE} Saha-Boltzmann equilibrium, and the other based on a {\em nebular} model for recombination.

Given $\Phi(T)$, we can write the line-center opacity factor $C_{\rm o} \equiv \kappa_{\rm o}/\rho$ as
\beq
C_{\rm o} = \frac{ \Phi(T)  f_{23} \sigma_{\rm cl} \lambda_{\rm o}}{\mu_{\rm e} \mu_{\rm p} m_{\rm p}^2 v_{\rm th}} 
\, ,
\label{eq:Co}
\eeq
where $\mu_{\rm p} \equiv \rho / \mu_{\rm p} m_{\rm p} = X =0.72$ is the Hydrogen mass fraction, and $\mu_{\rm e} \equiv  \rho/n_{\rm e} m_{\rm p} = 2/(1+X) = 1.16$ is  the mean molecular weight per electron for a fully ionised mix of H and He.
Here we have assumed for convenience a simple box profile $\phi_{\lambda_{\rm o}} = 1/\Delta \lambda_{\rm D}$ 
with a Doppler width $\Delta \lambda_{\rm D} = \lambda_{\rm o} v_{\rm th}/c$ associated with Hydrogen thermal speed $v_{\rm th} \equiv \sqrt{2kT/m_{\rm p}}$.

The LTE model is discussed in section 9 of the standard radiative transfer text by \citet{Hubeny14}.
From a Saha-Boltzmann analysis, their equation (9.5) gives
\beqa
\Phi_{LTE} (T) &=& C  ( g_2/g_1^+ )T^{-3/2} \exp (( \chi_I - E_2)/kT)
\\
&=&  1.66 \times 10^{-21} {\rm cm}^{3} \, \frac{\exp (3.945/T_4)}{T_4^{3/2}}
\label{eq:PhiLTE}
\ ,
\eeqa
where $\chi_I - E_2 =  3.4 $\,eV is the difference between the ionization energy and the excitation energy for $n=2$, 
and  $g_2/g_1^+ = 8$ is the ratio of the associated statistical weights.
The latter equality in (\ref{eq:PhiLTE}) gives numerical evaluation in terms of the scaled temperature $T_4 \equiv T/10^4 K$.

In the nebular recombination model, the equilibrium population in level $n=2$ depends on the ratio of recombination rate $\alpha_2 (T)$ into that level, to the spontaneous decay rate $A_{21}$ ($=9.466 \times 10^8$\,s$^{-1}$),
\beq
\Phi_{\rm neb} (T) = \frac{\alpha_2 (T)}{A_{21}}
\, .
\label{eq:Phineb}
\eeq
\citet{Pequignot91} give for the  recombination rate to level $n=2$ of Hydrogen,
\beq
\alpha_2 (T) = 
4.31 \times 10^{-13} \, \frac{\rm cm^3}{\rm s} \, \frac{T_4^{-0.62 }}{1+0.67 \, T_4^{0.64}}
\, .
\label{eq:alpha2}
\eeq

\begin{figure}
\begin{center}
\includegraphics[scale=0.45]{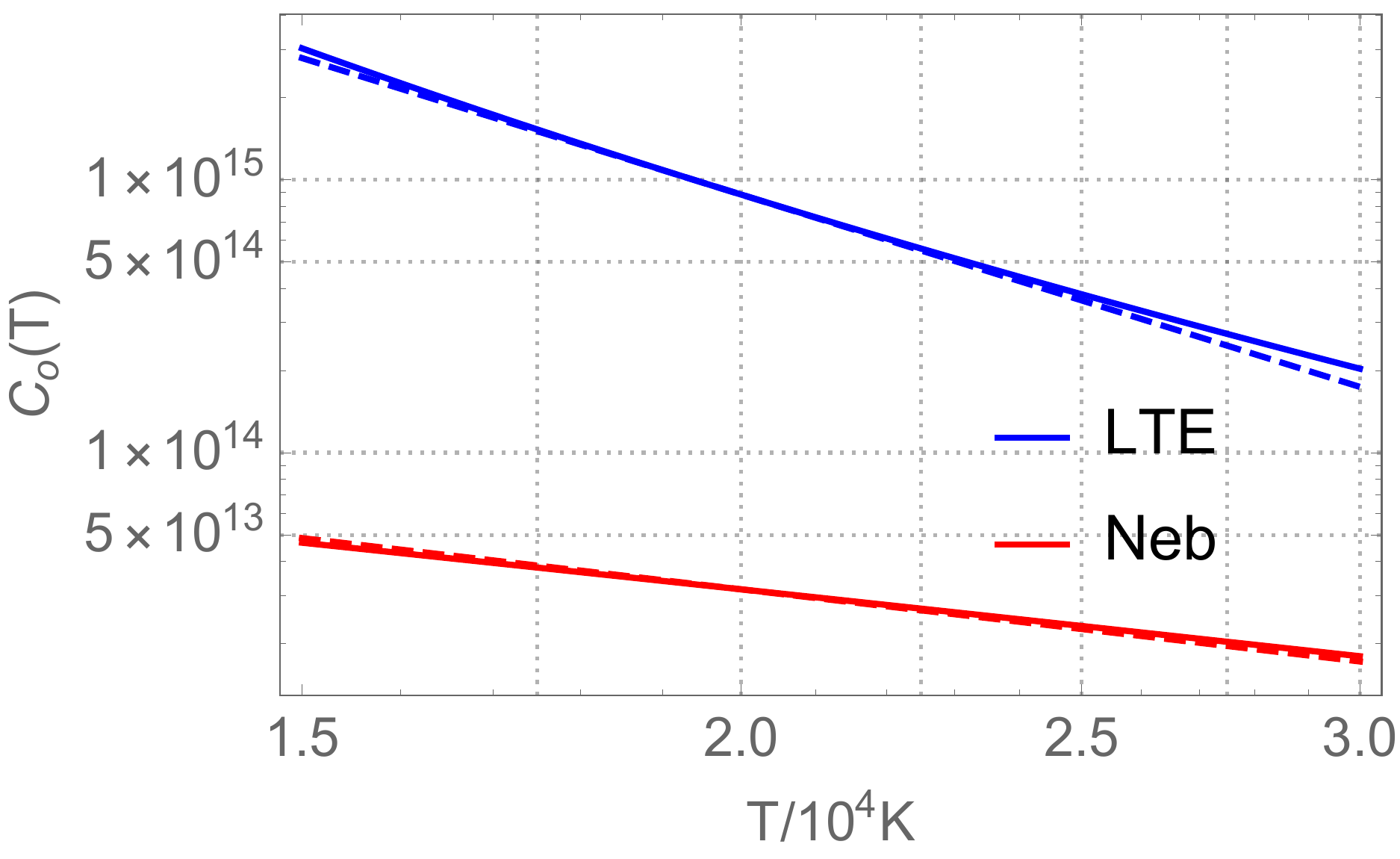}
\caption{
Log-log plot of the temperature variation of the opacity coefficient $C_{\rm o}$ (in CGS units, cm$^5$\,g$^{-2}$) for the LTE (Blue) and nebular (Red) models. The corresponding dashed lines show respective power-law fits, $C_{\rm o} \sim T^{-4}$ and $C_{\rm o} \sim T^{-1.5}$.}
\label{fig:figA1}
\end{center}
\end{figure}

By applying equations (\ref{eq:PhiLTE}) or (\ref{eq:Phineb}) into (\ref{eq:Co}),
figure \ref{fig:figA1} shows a log-log plot of  the variation of 
$C_{\rm o} (T)$
over the relevant early B-star temperature range $T_4 =1.5$ to 3, for both the nebular (red) and LTE (blue) models.
The corresponding power-law fits (dashed lines) quantify the weaker temperature decline (power index -1.5) for the nebular model compared to that  (power index -4) for  the LTE model.
Note also that the LTE values for $C_{\rm o}$ are much higher than those for the nebular model, with a ratio of 28, or about 1.45 dex, at a typical temperature $T_4=2$ in the midde of the early B-star range.

Finally, applying the  $C_{\rm o}$ equation (\ref{eq:Co}) into the $B_{\rm K1}$ equation (\ref{eq:BK1a}) then forms the basis for  the plots in figure \ref{fig:fig3} of the temperature dependence of $B_{\rm K1}$, again for both the LTE and nebular models.
The lower nebular value for $C_{\rm o}$ gives a higher $B_{\rm K1}$; 
but because of the weak scaling $B_{\rm K1} \sim \tau_{\rm K}^{1/4} \sim C_{\rm o}^{-1/4}$, when this lower value is combined with its lower power index, the net effect is to make $B_{\rm K1} (T)$ roughly {\em linear},
with the nebular values just offset by a roughly constant difference $\Delta B_{\rm K1} = + 90$\,G above the likewise linear relation for $B_{\rm K1} (T)$ from the LTE model.

\end{document}